\documentclass{aa}  

\usepackage{graphicx,textcomp}
\usepackage[varg]{txfonts}
\usepackage{hyperref}

\usepackage{graphicx}   
\usepackage{amsmath}    
\usepackage{booktabs} 
\usepackage{siunitx}
\usepackage{placeins}
\usepackage{tabularx}
\usepackage{array}

\begin{document}

\title{DBNets: A publicly available deep learning tool to measure the masses of young planets in dusty protoplanetary discs}

\author{A. Ruzza\inst{1}, G. Lodato\inst{1}  \and G. P. Rosotti\inst{1}}

\institute{\inst{1}Università degli Studi di Milano, Dipartimento di Fisica, via Celoria 16, 20133 Milano, Italy. \\
\email{alessandro.ruzza@unimi.it}}

\date{Received October 30, 2023; accepted February 08, 2024}
\titlerunning{DBNets}
\authorrunning{A. Ruzza et al.}

\abstract
{Current methods to characterize embedded planets in protoplanetary disc observations are severely limited either in their ability to fully account for the observed complex physics or in their computational
and time costs. To address this shortcoming, we developed \emph{DBNets}: a deep learning tool, based on convolutional neural networks, that analyses
substructures observed in the dust continuum emission of protoplanetary discs to quickly infer the mass of allegedly embedded planets.
We focussed on developing a method to reliably quantify not only the planet mass, but also the associated uncertainty introduced by our modelling and adopted techniques.
Our tests gave promising results achieving an 87\% reduction of the $\log M_p$ mean squared error with respect to an analytical formula fitted on the same data (\emph{DBNets} metrics: lmse 0.016, r2-score 97\%). With the goal of providing the final
user of \emph{DBNets} with all the tools needed to interpret their measurements
and decide on their significance, we extensively tested our tool on out-of-distribution data. We found that \emph{DBNets} can identify inputs strongly
outside its training scope returning an uncertainty above a specific threshold and we
thus provided a rejection criterion that helps determine the significance of the results obtained. Additionally, we outlined some limitations of our tool: it can be reliably applied only
on discs observed with inclinations below approximately 60°, in the optically thin regime, with a resolution $\sim 8$ times better than the gap radial location and with a signal-to-noise ratio higher than approximately ten.
Finally, we applied \emph{DBNets} to 33 actual observations of protoplanetary discs measuring the mass of 48 proposed planets and comparing our results with the available literature. We confirmed that most of the observed gaps imply planets in the sub-Jupiter~regime. 
\emph{DBNets} is publicly available at \href{http://dbnets.fisica.unimi.it}{dbnets.fisica.unimi.it}.}

\maketitle


\section{Introduction}
With the advent of astronomical interferometers such as the Atacama Large Millimeter/submillimeter Array (ALMA), we are now able to obtain spatially resolved observations of protoplanetary discs \citep{ALMAPartnership2015TheRegion}.
Over the past few decades, high-resolution data of these systems have revolutionized the field, enabling access to their complex morphology. This revealed the ubiquitousness of substructures such as gaps, rings, and spirals (e.g. \citealt{Andrews2018TheOverview, Isella2016RingedALMA, vanTerwisga2018V1094Star, Dipierro2018RingsALMA, Long2018GapsRegion, Fedele2018ALMA209,Huang2020Large-scaleLup, Cieza2021TheResolution}; see also \citealt{Bae2023StructuredDisks} review).

A plethora of different mechanisms have been proposed to explain and model these substructures, including several hydrodynamical \citep{Dullemond2018Dust-drivenDisks, Takahashi2014TWO-COMPONENTSTRUCTURES, Barge2017GapsDust} and magnetohydrodynamical \citep{Hawley2001GlobalDisks, Ruge2016GapsGrains, Takahashi2018StructureWind} processes, self-induced dust pile-ups \citep{Gonzalez2015ALMAPlanets}, and condensation processes at snow lines \citep{Zhang2015EVIDENCEDISK, Pinilla2017DustDisks}.
However, another possible explanation is the presence of embedded planets tidally interacting with the disc material \citep{Zhang2018TheInterpretation, Rosotti2016TheObservations, Dipierro2015OnTau}. Under this assumption, the observation and modelling of these signatures enable the detection and characterization of a population of young planets hard to access with commonly used methods of exoplanet search. Substructures observed in the dust millimetric continuum emission are especially suited for this purpose. In fact, although there are other direct (e.g. \citealt{Keppler2018Discovery70, Currie2022ImagesAurigae, Reggiani2014DISCOVERYFORMATION}) and indirect (e.g. \citealt{Pinte2018KinematicDisk, Casassus2019Kinematic100546, Calcino2022MappingKinematics}) methods to detect planets in protoplanetary discs, these are mostly sensitive to super-Jupiter planets while, with the currently achievable sensitivity, even planets up to a few Earth masses can produce visible signatures in the dust distribution \citep{Zhu2014PARTICLEDISKS, Rosotti2016TheObservations, Dipierro2017AnDiscs, Dong2018MultipleSystems}.

From numerical simulations, some empirical formulae linking gaps' widths and depths with planet masses have been proposed \citep{Rosotti2016TheObservations, Dipierro2017AnDiscs,  Lodato2019TheDiscs}, but they suffer from several limitations connected to the modelling assumptions. Furthermore, they are not able to fully take some complex features of observations into account, such as multiple or asymmetric substructures, and the proposed formulae often depend on the chosen definition for the gap properties. A more
accurate analysis requires time-consuming numerical modelling with a fine-tuning of the disc and planets’ properties with the goal of reproducing the observed features (e.g. \citealt{Fedele2018ALMA209, Toci2019Long-lived169142, Veronesi2020IsPlanet}). The high computational cost of this approach limits its applicability to large surveys (e.g. \citealt{Andrews2018TheOverview, Long2018GapsRegion, Cieza2021TheResolution}) and prevents the use
of statistically robust methods for parameter inference resulting in the lack of a proper way to determine their statistical significance.

Machine learning techniques can address the outlined shortcomings and provide an improved method for these analyses. The high flexibility of deep neural
networks can be used to efficiently capture, from a broad dataset of synthetic observations, any correlation between planet properties and substructure morphologies exploiting the entire observation, including asymmetric and not immediately visible features. Once trained, the application of a neural network to new data takes less than a second, allowing the scaling of our approach to the analysis of arbitrarily large datasets.

Recently, several works have started researching machine learning methods to model and interpret protoplanetary discs' observations, focussing on edge-on discs \citep{Telkamp2022ADisks}, discs' spectral energy distributions \citep{Kaeufer2023AnalysingLearning}, planet-induced substructures in the gas \citep{Mao2023PPDONet:Systems}, and kinematical signatures of embedded planets \citep{ Terry2022LocatingLearning}. Additionally, the same issue of measuring the mass of gap-opening planets from dust substructures has also been addressed \citep{Auddy2020ADisks, Auddy2021DPNNet-2.0Gaps, Auddy2022UsingDisks, Zhang2022PGNets:Discs}.
\cite{Auddy2020ADisks} implemented a feed-forward neural network that estimates the planet mass from a manually measured gap width and, optionally, some disc properties. However, this approach is still limited by the use of a single property of disc gaps instead of their entire morphology. \cite{Auddy2021DPNNet-2.0Gaps} and \cite{Zhang2022PGNets:Discs} overcame this limitation using convolutional neural networks (CNNs) that directly take a dust continuum observation or density map as input.
Still, two major issues were left partly (or completely) unaddressed: the lack of a proper
way to quantify the uncertainties of their estimates and the sensitivity of the final models to non-idealities not included in the training data. 
\cite{Auddy2022UsingDisks} addressed the former, implementing Bayesian neural networks but they went back to feed-forward architectures using manually measured gap features instead of the whole disc observations.
Another issue was pointed out by \cite{Zhang2022PGNets:Discs} who observed that different instances
of the same neural network architecture, trained on the same data but with different random seeds
for data shuffling and weight initialization, resulted in trained models that, despite having
comparable statistical performances on the whole test set,
often produced very different mass predictions when compared
on single discs. 

In this work, we present the tool \emph{DBNets} (short for \mbox{DUSTBUSTERS} Nets) which analyses substructures observed in the dust continuum
emission of protoplanetary discs to quickly infer the mass of allegedly embedded planets. We used CNNs similarly to \cite{Auddy2021DPNNet-2.0Gaps} and \cite{Zhang2022PGNets:Discs}, but we specifically address the aforementioned
problems of uncertainty quantification and high variance of the results by developing and testing an ensemble technique inspired by previous works
in separate fields \citep{Lakshminarayanan2016SimpleEnsembles, Cobb2019AnRetrieval}. Additionally, we extensively tested the robustness of our tool and constrained its scope of applicability by evaluating the effect of different phenomena that may lead to incorrect
mass estimates. 
To train and test our deep learning models, we generated a dataset of synthetic observations using hydrodynamical simulations and analytical calculations to compute the disc brightness temperature.
In doing this, we explored a slightly larger parameter space and adopted different choices with respect to the work of  \cite{Auddy2021DPNNet-2.0Gaps} and \cite{Zhang2022PGNets:Discs}, which would make it interesting to test our models against their datasets.
Finally, we applied \emph{DBNets} to a large sample of real observations providing the most comprehensive analysis of protoplanetary disc's gaps for the measurement of putative planets' masses. We compared and discussed the results obtained with those currently
available in literature.

This paper is organized as follows.
In Sect. \ref{sec:methods} we present our methodology presenting the code and models that we used to generate synthetic data and the deep learning techniques that we implemented.
We present our results in Sect. \ref{sec:results}, discussing the dataset that we generated and the performance of the CNNs' ensemble on our test set.
In Sect. \ref{sec:real} we explain how we applied \emph{DBNets} to actual observations of 33 unique protoplanetary discs for a total of 48 gaps that are all assumed to host a planet. We then discuss our results and their deviation from the available literature.
In Sect. \ref{sec:discussion} we first compare \emph{DBNets} with a simpler empirical relation assessing the better performance of our tool. We then discuss the uncertainty quantification method that we implemented and our testing of \emph{DBNets} against out-of-distribution data. In the same section, we also discuss the main caveats of our tool in the current version.
We provide our conclusions and main plans for future developments in Sect. \ref{sec:conclusions}.

\section{Methods}
\label{sec:methods}
We aim to use a machine learning method to link the morphology of protoplanetary disc substructures observed in the dust continuum emission to the mass of possibly embedded planets.
For this purpose, we implemented the tool \emph{DBNets} based on CNNs: deep learning models specifically designed to process images. Through an iterative process, called training, CNN's parameters are optimized to fit an unknown relation between features of a large dataset. The optimization is carried out by minimizing a loss function that measures the difference between the model's outputs and target values. The process thus requires a sample of labelled data, called training set, in which target values are known.

\begin{figure}[t]
    \centering
    \includegraphics[width=\linewidth]{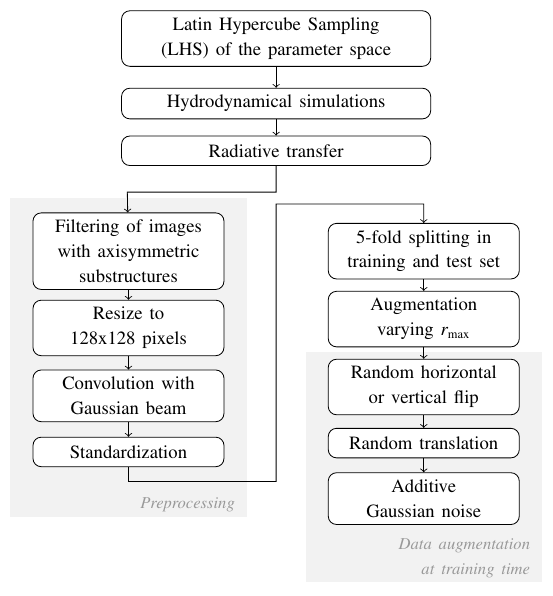}
    \caption{Pipeline adopted for the generation and preprocessing of the mock observations that we used to train and test our machine learning models. The $r_\text{max}$ denotes the outer radius of the simulated discs.}
    \label{fig:preprocess}
\end{figure}

As the first step of our work, we assembled a dataset of synthetic images of protoplanetary discs' dust emission running hydrodynamical simulations and post-processing the results to include some features of actual resolved observations. Figure \ref{fig:preprocess} summarizes this image generation and preprocessing procedure outlining all its steps. We then used part of the obtained sample of images to train a CNN with the goal of analysing a disc continuum observation to estimate the mass of an allegedly embedded planet. 

One of the main challenges of deep learning techniques is quantifying the uncertainty of the measures obtained with these tools. The inclusion of an effective and reliable method for this purpose is of paramount importance in delivering a tool deployable in real-case scenarios to obtain significant results able to sustain further discussions and statistical analysis.
To address this issue, we implemented an ensemble approach, which we thoroughly explain in the next sections, able to quantify both the uncertainty due to physical degeneracies in the planet-generated substructures and the uncertainty introduced by our particular choice of deep learning model.
Our implementation also allowed us to evaluate the impact of phenomena that we did not include in our set of synthetic data but might affect actual observations. We thus constrained the scope of applicability of our tool providing a clear outline of the assumptions under which one should interpret its measures.

The next subsection describes the hydrodynamical simulations that we ran explaining their numerical setup and the dynamical parameters that we chose to sample. In Sect. \ref{sec:post} we explain the additional post-processing steps applied to the simulations' results. Finally, in Sects. \ref{sec:cnnarch} and \ref{sec:ensemble}, we introduce the machine learning methods used: the CNNs' setup and the details of the ensembling technique.

\subsection{Simulations}
We used the Eulerian mesh code \textsc{fargo3d} \citep[][]{Benitez-Llambay2016FARGO3D:CODE} to run 2D  hydrodynamical simulations of protoplanetary discs with two fluid components modelling gas and dust \citep{Benitez-Llambay2019AsymptoticallyFARGO3D}. We used a locally isothermal equation of state for the gas component and assumed vertical hydrostatical equilibrium to prescribe the temperature profile through the disc scale height.
Solids are modelled as one population of pressureless dust experiencing a drag force with a fixed Stokes number. Additionally, mass diffusion, with a Schmidt number (Sc) of 1, is added to the dust density \citep{Weber2019PredictingDisks}.
We neglected the disc self-gravity and the dust feedback on the gas component to simplify our numerical model and reduce the parameter space.
In each simulation, a non-accreting and non-migrating planet is placed at $r_0=1$ (code units) on a circular orbit from the beginning. We varied the planet/star mass ratio and 5 disc properties in the different simulations we ran.

\subsubsection{Numerical setup}

\begin{table}
        \centering
         \caption{Common setup of the hydrodynamical simulations.}
 \label{tab:num_setup}
        \begin{tabular}{l|c} 
        \textbf{Dimensions} & 2D \\
        \textbf{Equation of state} & locally isothermal \\
        \textbf{Fluids} & 2 (gas+dust) \\
        \textbf{Planet position} & 1 (code units) \\
        \textbf{Mesh type} & cylindrical, log spaced in $r$ \\
        \textbf{Radial domain} & [0.4, $r_\text{max}$] ($r_\text{max}=3$ or $5$) \\
        \textbf{Resolution $(\xi \equiv l_\text{cell}/H)$} & $\leq$0.2 \\
        \textbf{Simulated time (planet orbits)} & 1500 \\
        \textbf{Boundary conditions} & open, wave killing \\
        \textbf{Dust diffusion} & yes \\
        \textbf{Schmidt number (Sc)} & 1 \\
        \textbf{Planet migration} & no \\
        \textbf{Dust feedback} & no \\
        \textbf{Disc self-gravity} & no \\
        \end{tabular}

\end{table}

The domain of the simulations consists of a 2D mesh in cylindrical coordinates ($\phi, r$) with $\phi \in [0;2\pi)$ and $r\in [0.4, r_{\text{max}}]$. All physical quantities are given in code units defined such as $G=r_0=M_{\star}=1$. We set the upper end of the radial domain to $r_{\text{max}}=3$ for simulations with a  planet-star mass ratio lower than $5\cdot10^{-3}$, and to $r_{\text{max}}~=~5$ for more massive planets expected to generate wider substructures. With this setting, we checked that the disc morphology is not affected by the outer boundary. The mesh is split uniformly in $\phi$ and $\log r$ to obtain a grid of $N_\phi \times N_{r}$ nearly squared cells. 
We measure the numerical resolution of our simulations by comparing the dimension $l_\text{cell}$ of these approximately squared cells with the vertical scale height ($H$) at their position. The cell dimensions are set for each simulation to obtain a resolution of $\xi =l_{\text{cell}}/H= 0.2$ at the planet location which we found to be an adequate compromise between computational costs and numerical accuracy. We observed however that a $N_\phi$ greater than approximately 600 is required to avoid computational artefacts due to the low angular resolution. Hence, in simulations of thicker discs, where the $\xi=0.2$ prescription would have resulted in a lower resolution, we instead forced a minimal grid dimension of $600 \times 193$.
We ran the simulations for 1500 orbits of the planet, equivalent to approximately $5.3\times 10^5$ yr for a planet at $50$ au orbiting a 1 M$_\odot$ star. Table \ref{tab:num_setup} summarizes the fixed setup of all the hydrodynamical simulations that we ran.

Initial and boundary conditions were set with the goal of modelling an annulus, around a planet, of a viscously accreting disc (e.g. \citealt{Armitage2020AstrophysicsEdition}). We consider the steady state solution of the evolution equation, with a constant accretion rate ($\dot M$), to set the initial conditions
\begin{align}
    \dot M &= 2\pi v_r r \Sigma(r),
     \\
    \Sigma(r) &= \frac{\dot M}{3\pi\nu}\left(1-\sqrt{\frac{r_{\text{in}}}{r}} \right) \sim \frac{\dot M}{3\pi\nu},\label{equ:dens_steady_state}
\end{align}
where $v_r$ is the radial velocity of the fluid and $\Sigma$ the initially unperturbed and axisymmetric disc surface density.
For the disc viscosity $\nu$, we used the \cite{Shakura1973BlackAppearances} prescription setting $\nu= \alpha c_s H$ with a constant $\alpha$ for each disc. With $c_s$ we denote the fluid sound speed.
The radial profile of the disc aspect ratio $h=H/r$ is set to a power law whose index ($\beta$) is called flaring index and controls the disc temperature profile under the assumption of vertical hydrostatical equilibrium
\begin{equation}
    \frac H r = \frac {c_s} {\Omega r} = h(r) = h_0 \left(\frac{r}{r_0} \right)^{\beta}, \label{eq:ar}
\end{equation}
with $\Omega$ being the fluid angular velocity.
This setup prescribes the following profile for the radial velocity $v_r$
\begin{equation}
    v_r = \frac 3 2 \alpha H^2 \Omega /r =  \frac 3 2 \alpha h_0^2 \left(\frac r {r_0} \right)^{2\beta -0.5} r_0 \Omega(r_0)
\end{equation}
which we used to initialize the simulations.
Gas and dust initial surface density profiles are set to a power law
\begin{equation}
    \Sigma(r)  =\Sigma_0 \left( \frac r {r_0}\right)^{-\sigma}, \label{eq:sigma0}
\end{equation}
with $\Sigma_{0, \text{gas}}=10^{-3}$ for the gas component and $\Sigma_{0, \text{dust}}=10^{-5}$ for the dust. Since we neglect self-gravity and dust feedback, the results of the hydrodynamic simulations do not depend on these parameters under appropriate rescaling.

We used open boundary conditions, copying values from the active cells to the ghost cells, for both the density and radial velocity to allow the disc to evolve freely. We prefer this solution to the alternative of fixing the inner density with a closed boundary because, when a dust trap forms, it allows the formation of a central hole.
Additionally, we applied wave-killing boundary conditions to the density and velocity of the gas component, similarly to \cite{DeVal-Borro2006AInteraction}, relaxing their values to the azimuthal average. 

\subsubsection{Parameter space sampled}

A summary of the parameters explored in our dataset of simulations is given in Table~\ref{tab:par_table}. We used Latin hypercube sampling (LHS) to draw 1000 samples in the space of the 5 parameters with a uniform distribution in the given ranges of values. For parameters spanning several orders of magnitude, we sampled their log values instead.
The disc flaring index ($\beta$) and the slope of the surface density profile ($\sigma$) were not extracted independently. Instead, we sampled the flaring index and then set $\sigma$ according to the viscously accreting steady state solution given by Eq.~(\ref{equ:dens_steady_state}) combined with the $\alpha$-prescription (assuming a constant $\alpha$) and Eq.~(\ref{eq:ar}) to obtain the relation
\begin{equation}
    \label{equ:slopef}
    \sigma = 2\beta + 1/2.
\end{equation}
The explored range of planet masses varies approximately between $3 \text{M}_{\oplus}$ and $10 \text M _{\text{J}}$. We are especially interested in the lower end of this range because dust substructures may be the only way to detect and characterize these planets due to the weakness of the perturbations expected in the gas component. 
Stokes number's values correspond approximately to grain sizes between $10$~and~$10^{-2}$~mm for a disc surface density $\Sigma_0 \sim \SI{3.5}{\g\per\square\cm} \left({M_{\text{gas}}}/{M_{\odot}}\right)\left( {r_0}/{50 \text{au}}\right)^{-2}$.
We used a fixed Stokes number for the entire disc which implies that its conversion to the dust grain size results in a map of values varying with $r$ according to the gas surface density profile.

\begin{table}
        \centering
        \caption{Dynamical properties sampled in the simulation's dataset.}
 \label{tab:par_table}
        \begin{tabular}{lccr} 
                \toprule
                Property & Symbol & Values & Type of sampling\\
                \midrule
                $\alpha$-viscosity & $\alpha$ & $10^{-4} - 10^{-2}$ & log \\
                Stokes number & $St$ & $10^{-3} - 10^{-1}$ & log\\
                Aspect ratio & $h_0$ & $0.03 - 0.1$ & lin \\
        \begin{tabular}[x]{@{}c@{}} Planet/star \\ mass ratio \\ \end{tabular} & $M_p$ & $10^{-5} - 10^{-3}$ & log \\
        Flaring index & $\beta$ & $0 - 0.35$ & lin \\
        Sigma slope & $\sigma$& $0.5 - 1.2$ & Eq. (\ref{equ:slopef})\\
                \bottomrule
        \end{tabular}
 
\end{table}

\subsubsection{Image generation}

We collected in our dataset the dust density maps obtained from the hydrodynamical simulations after 500, 1000 and 1500 planetary orbits. From the density, we computed the expected dust continuum emission at wavelength $\lambda=\SI{1.3}{\mm}$ producing mock observations that we used to train and test the machine learning models.
We assumed emission in the Rayleigh-Jeans regime, computing the disc brightness temperature according to
\begin{equation}
    T_s = T_d\left(1-e^{-\tau}\right), \label{eq:brightT}
\end{equation}
where $\tau$ denotes the dust optical depth.
We set the disc temperature $T_d$ consistent to the hydrodynamical simulations obtaining it from the disc aspect ratio since $H/r =c_s/v_k \propto \sqrt{T r}$. For each simulation, this results in 
\begin{equation}
    T_d \propto \left(\frac{r}{r_0}\right)^{2\beta -1}. \label{eq:temp}
\end{equation}
Multiplicative constants can be neglected because intensity maps are then standardized.

The optical depth $\tau$ is computed from the dust surface density ($\Sigma_\text{dust}$) as
\begin{equation}
    \tau(r, \phi) \sim \Sigma_{\text{dust}}(r, \phi)\kappa[a_\text{max}(r, \phi), \lambda], \label{eq:od}
\end{equation}
where $\kappa$ is the opacity of dust grains that depends on their size $a$ and wavelength $\lambda$ of the radiation.
We need to remember that the assumption of fixed $St$ in our simulations implies that the dust size changes at each location in the disc, depending on the local gas properties. So, we first obtained a map of the grain sizes from the assumption of fixed Stokes number across the disc. We assumed Epstein drag and computed the grain size in each cell through the equation
\begin{equation}
    a_\text{max} =  \frac2{\pi}\frac{St \cdot \Sigma_{0, \text{gas}}}{\rho_{\text{grain}}} \left(\frac r {r_0} \right)^{-\sigma}
\end{equation}
in which we used the initial gas surface density profile given by Eq. (\ref{eq:sigma0}) under the assumption that it gets negligibly perturbed by the planet and we set the intrinsic dust grain density to $\rho_{\text{grain}} = 3$~\si{\g\per\cubic\cm}.
Next, we assigned to each cell the dust opacity computed with \cite{Birnstiel2018TheModel} for $\lambda=1.3$~\si{\mm} and averaged for a population of particles of different grain sizes modelled with the power law $n(a) \propto a^{-3.5}$ with a cut-off set at $a_{\text{max}}$ equal to the value previously computed from the Stokes number. We used \cite{Birnstiel2018TheModel} default composition that assumes dust grains of water ice, refractory organics, troilite and astronomical silicates
using mass and volume fractions from the available literature.
At this point, we are thus able to compute the optical depth map with Eq. (\ref{eq:od}) using the dust density obtained from the hydrodynamical simulations.

While the results of the hydrodynamical simulations are invariant to a rescaling of the surface density, the same is not true for the results obtained after this step of our procedure. We assumed, in code units, $\Sigma_{0, \text{gas}} = 10^{-3}$ and $\Sigma_{0, \text{dust}} = 10^{-5}$ that are equivalent to discs with gas masses between 0.02 and 0.03 M$_\odot$ in the simulated region, assuming $M_\star = 1$~M$_\odot$ and $r_0=50$~au. It is important to note, however, that they could be easily changed or sampled by repeating the above procedure without significant computational effort.

We are expecting to observe two main differences between the density and intensity maps that justify our decision of carrying out this additional step.
First, overdense regions where dust piles up can become optically thick, reducing the contrast between those pixels and the rest of the image with respect to the density map of the same disc.
Additionally, due to the temperature profile (Eq. \ref{eq:temp}), outer regions are expected to appear darker in intensity maps.

Real observations are affected by additional details such as the observational setup and \textsc{clean} methods used to obtain images from visibilities. We neglected these effects to avoid the introduction of additional parameters and because they are strongly dependent on the details of each specific observation.
The only non-ideality that we introduced in our mock observations is the finite resolution which we simulated convolving the intensity maps with a Gaussian beam. We used the same beam size ($\sigma \sim 0.12 r_0$) for the whole dataset choosing this value in consideration of available protoplanetary discs' data. We looked at the same sample of observations discussed in Sect. \ref{sec:real} and selected the mean of their resolution (see Appendix \ref{app:resreal}).
At the end of this step, we thus have synthetic images in the form of beam-convolved brightness temperature maps.

\subsection{Cross validation}
\label{sec:cross}
We divided the synthetic observations in our dataset in training and test sets implementing a 5-fold cross-validation scheme. 
At this stage, we decided to implement a training-test split rather than a training-test-validation split due to the small size of the original dataset. For our methodology, the additional validation set is not strictly necessary as we did not perform any hyperparameter optimization based on the results obtained on the test set. We manually tuned, by trial and error, some of the models' hyperparameters but all of our choices relied only on the results obtained on the training set. We thus did not introduce any bias requiring an additional set of data to independently asses our tool's performance.
In future developments, an automatic hyperparameter tuning may be attempted to improve the tool's performance. This will however require to increase the dataset's size to allow the presence of a separate validation set.
In the following sections, when showing or referring to applications of the model to the test set, we combine the results obtained for each fold using the models trained and tested on the respective separate sets.

\subsection{Data preprocessing and augmentation}
\label{sec:post}

On the synthetic images, we carried out additional preprocessing, data augmentation and filtering. The aim of these additional steps is to help the training of the neural networks, avoid overfitting and improve generalization capabilities.
We used different radial domains for the simulation with more massive planets to avoid artefacts related to the boundary but we then cropped all the results for $r \in [0.4, 4]$.
Pixel values for each image are the features that the neural network will learn to link to the planet's mass. Increasing the number of pixels can thus result in an improvement of the predictions but, at the same time, increases the computational complexity of the training procedure. Hence, we downsampled each image to the size of $128 \times 128$ pixels which we found to be the best solution for this trade-off. We used the \textsc{inter\_area} interpolation method implemented in the python package \textsc{opencv} to apply the resize. This algorithm computes the value of each pixel in the shrunk image as a weighted average of corresponding pixels of the original image in a selected box whose size depends on the ratio between the original and target dimensions. Each pixel is weighted with the area that it occupies in this box.
After this resizing, each image is standardized by subtracting the mean value of its pixels and dividing by their standard deviation.

The aim of our work is to extract information on the planet's mass from the morphologies of substructures produced by the planet-disc interaction. The tool we are building will not be used on smooth images without gaps, rings or other visible axisymmetric substructures. For this reason, we also removed these substructure-free images from the training set identifying them by looking at the logarithmic derivative of the azimuthally averaged dust density in an annulus near the planet. We assumed that a positive value of the derivative indicates the presence of an axisymmetric substructure.

Initially all the discs in the training set extend up to the boundary of the image, which corresponds to $r=4$ in code units. However, real images of discs can have different sizes and, in some cases, their rescaling to match the conditions of the training set results in discs smaller than the image. We found this situation to be problematic leading the CNN to interpret the depleted region outside the disc as a gap.
To prevent this behaviour and guide the neural networks to fit the substructures we augmented the training dataset with the following procedure. For each image, we randomly extracted 3 values for $r_{\text{max}}$ in uniformly spaced intervals between 2 and 4. For each one of these values, we generated a copy of the selected image where the disc extends up to the respective $r_{\text{max}}$. We extrapolated the surface density using the initial condition power law when the extracted value of $r_{\text{max}}$ is larger than the radial domain of the simulation.
This step triplicates the size of the dataset. Note in Fig. \ref{fig:preprocess} the precise order in which we carried out all the explained steps. In particular, note that each process of data augmentation is carried out after splitting the data in training and test sets as explained in Sect. \ref{sec:cross} and is applied only to the training set images. For each image of the test set instead, only one value for $r_\text{max}$ is extracted.
During the CNNs' training, we perform additional data augmentation by randomly flipping each image horizontally or vertically to enforce rotation invariance, applying a random translation up to 10\% in each direction, and adding Gaussian noise with mean 0 and standard deviation 0.01.

\subsection{CNN architecture}
\label{sec:cnnarch}

\begin{figure*}[ht]
    \centering
    \includegraphics[width=\textwidth]{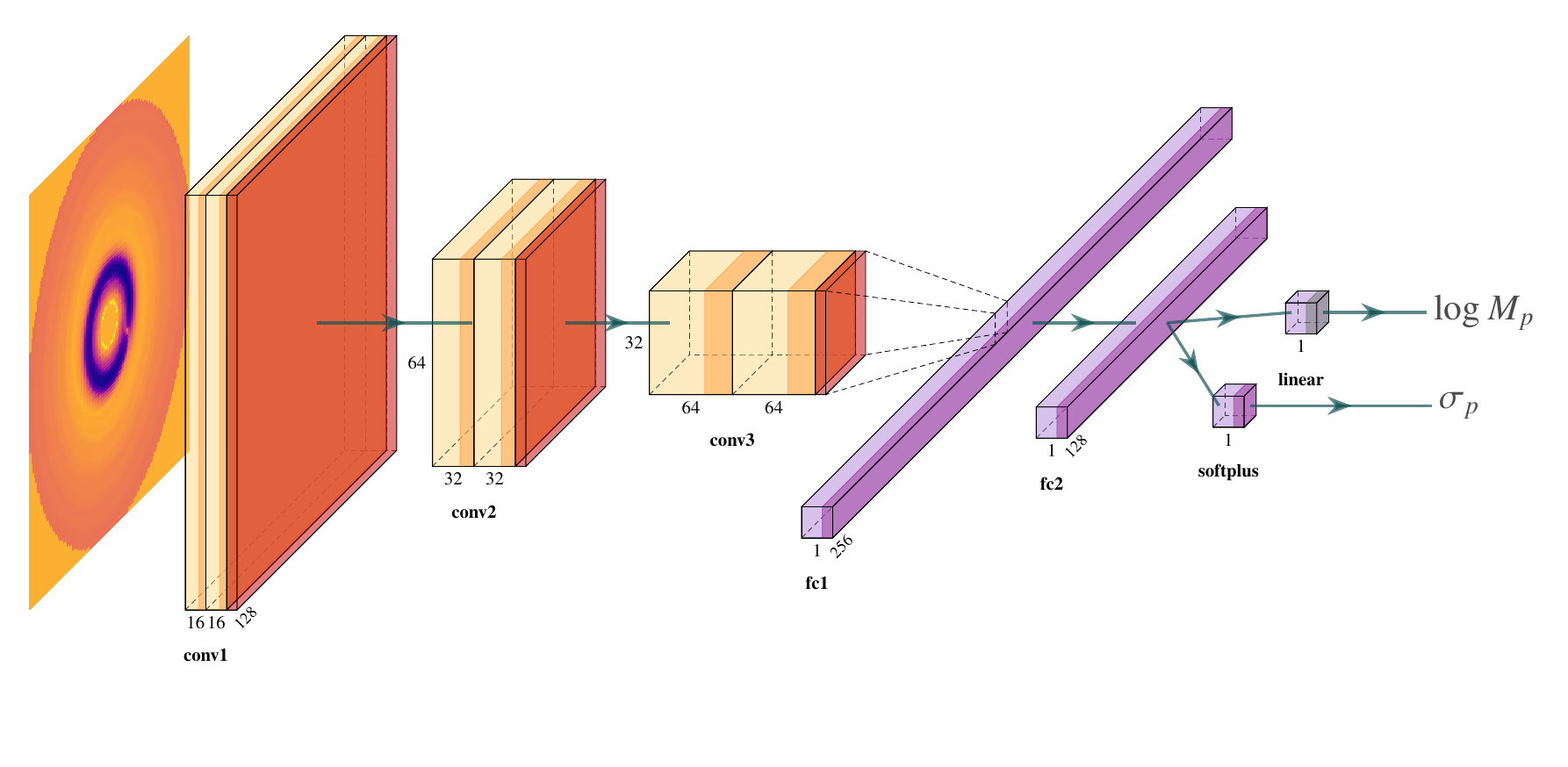} 
    \caption{Architecture of the trained CNNs. The disc image is fed to the convolutional part of the neural network made of three blocks. Each of them consists of two convolutional layers with filter size 3x3 (light orange) followed by a max pooling layer with filter size 2x2 (dark orange). The second part of the CNN consists of two fully connected layers. We omit in this image the dropout and normalization layers. }
    \label{fig:cnn}
\end{figure*}

We implemented a  CNN with \textsc{tensorflow} \citep{Abadi2015TensorFlow:Systems} using the architecture shown in Fig.~\ref{fig:cnn}. The peculiarity of CNNs with respect to regular neural networks lies in the introduction of convolutional layers designed for their application to data, such as images, characterized by high dimensionality and the importance of the relative position of their features. We constructed a relatively simple model that, inspired by affirmed architectures such as LeNet-5 \citep{LeCun1998Gradient-basedRecognition}, AlexNet \citep{Krizhevsky2012ImageNetNetworks} and VGG16 \citep{Simonyan2014VeryRecognition}, is formed by consecutive blocks of two convolutional, one pooling layer and a normalization layer. In each convolutional block, we increased the number of feature maps and reduced their size. The neural network is finally completed by two dense layers that transport the information up to the two outputs.
We also included dropout layers to improve training and generalization. We placed one of them before each dense layer using a dropout rate of 10\% that we kept unvaried among all the CNNs of our ensemble.
The activation function used in all the intermediate layers is the leaky ReLU that we chose instead of the ReLU to solve a problem of vanishing gradients that we encountered during our first attempts with the implementation.
To construct the ensemble we need to train over 50 instances of the CNN and we thus decided to avoid using very deep models such as ResNet50 to reduce the computational costs associated with their training.

The neural network takes in input a $128 \times 128 $ pixels image, whose values are expected to be standardized, and returns two real numbers $\log M_p$ and $\sigma_p$. The latter is always positive thanks to the \textsc{softplus} activation function applied before its output. We interpret them as the parameters of a Gaussian probability density function (pdf) for the logarithmic value of the mass of the planet supposedly embedded in the disc substructures $(M_\text{targ})$:
\begin{align}
    p(\log M_{\text{targ}} | x, \theta) &= \mathcal{N}(\log   M_p, \sigma_p) =\nonumber \\ &=\frac{1}{\sqrt{2\pi\sigma_p}}e^{-{(\log M_p - \log   M_{\text{targ}})^2}/{2\sigma^2_p}}. \label{eq:pdf}
\end{align}
Using Bayesian formalism we made explicit the dependence of this pdf on the input image ($x$) and the CNN parametrization ($\theta$). We call the standard deviation $\sigma_p$ of this predicted distribution `physics uncertainty' as it should represent the uncertainty on the mass prediction due to the physical degeneracies that we managed to represent in the training set.

To train the model we designed a loss function according to the maximum likelihood estimation principle obtaining the following expression that is minimized during the training phase:
\begin{equation}
    \mathcal{L} = \frac{1}{N}\sum_{i\in \mathcal{B}} \log\sigma_{p,i} + \frac{(\log   M_{p,i} - \log M_{\text{targ},i})^2}{2\sigma_{p,i}^2}, 
\end{equation}
where the sum runs over the entire batch ($\mathcal{B}$) of $N$ synthetic observations with target planet masses $M_{\text{targ}}$. 
We encountered some difficulties in directly trying to optimize this function such as failed or slow convergence. To address this problem we introduced a `warm-up training phase' prior to the actual training where we substituted the previously presented loss function with the simpler logarithmic mean squared error (lmse) of the mass prediction ignoring the second output of the CNN:
\begin{equation}
\text{lmse} = \frac{1}{N}\sum_{i\in \mathcal{B}} (\log   M_{p,i} - \log M_{\text{targ},i})^2. \label{eq:lmse}
\end{equation}
We used the \textsc{nadam} optimizer setting the learning rate to $10^{-4}$ and leaving all the other parameters to their default values. We used a small batch size of 16 elements.
Every CNN was trained for 300 epochs (plus 200 for the warm-up phase) implementing early stopping to arrest the optimization when the loss function evaluated on the test set reaches a plateau.

\subsection{Ensemble}
\label{sec:ensemble}
The complexity of deep learning models, such as the CNN we used, is reflected in the loss function that usually presents several local minima. The specific one reached during the training phase of the model can depend on several factors such as the CNN architecture, the shuffling of the training sample, the optimization algorithm or the weight initialization of the model. As a result, two models obtained from different training can return different predictions, an issue that was also reported by \cite{Zhang2022PGNets:Discs} in their work. We called this the `model uncertainty' of the prediction and is associated to the dependence on the parameters $\theta$ of the CNNs embedded in the returned probability density function Eq. (\ref{eq:pdf}).

Ensemble methods have been used \citep{Lakshminarayanan2016SimpleEnsembles, Cobb2019AnRetrieval} to mitigate this issue and offer an estimation of the prediction uncertainty. We implemented here a similar pipeline.
We trained 10 CNNs on each fold in which we divided the dataset. All the performance evaluations of our method on the test set only use the 10 models of the respective fold to avoid testing on data used for training. However, we deploy the ensemble of all 50 models for its application on real observations or for external testing.
The models differ by the random seed used for the CNN initialization and for the shuffling of the training data which \cite{Lakshminarayanan2016SimpleEnsembles} proved to suffice in providing enough variety to the ensemble.
As we anticipated in the previous section, each individual $i$ in the ensemble $\mathcal{E}$ returns its estimate of the planet's mass in the form of the probability density function Eq. (\ref{eq:pdf}).
We combine these pdfs  marginalizing over the $\theta$ dependence through
\begin{equation}
    p(\log M_p|x) \sim \sum_{i\in \mathcal{E}} p(\log M_p|x, \theta_i) \pi_i, \label{eq:ensres}
\end{equation}
where the weights $\pi_i$ can quantify the relative importance of the different parameterizations. In the tool we built, we implemented two possible ways of assigning the $\pi_i$ values corresponding to two different interpretations of the ensemble and its members.
The first one is called `peers mode' and is based on the assumption that all the individuals have reached the same level of knowledge of the topic. In this mode, the weights $\pi_i$ are set to
\begin{equation}
    \pi_i^{\text{peers}} = e^{-L_i},
\end{equation}
where $L_i$ is the value of the loss function obtained by the $i^{\text{th}}$ model during its training. With this definition, by design
\begin{equation}
    \pi_i^{\text{peers}} = p(\mathcal{D} | \theta).
\end{equation}
We are thus giving more weight to models that, if true, are more likely to reproduce the training set $\mathcal{D}$.
We also implemented the `experts mode' where we additionally multiplied each weight $\pi_i$ times the inverse square value of the physics uncertainty estimated by the respective neural network:
\begin{equation}
    \pi_i^{\text{experts}} = e^{-L_i} \frac{1}{\sigma_i^2}.
\end{equation}
This mode assumes that, in the ensemble, every single individual may have specialized in a diverse region of the parameters' space with the standard deviation of the predicted pdf quantifying the model's confidence in that specific prediction. 
We discuss the differences observed between the two modes in Appendix \ref{sec:peersvsphysics}. We anticipate however that on the test set they both provide very similar results. Unless otherwise stated, in the following sections we thus only discuss results obtained with the peers mode.

Following this procedure, the ensemble's output to a given disc image $x$ is the probability density function given by Eq. (\ref{eq:ensres}) that can be accurately studied and interpreted. For most of the analysis we carried out, we extracted from the pdf a prediction in the form
\begin{equation}
    (\log M_p )^{+\sigma_+}_{-\sigma_-}
\end{equation}
that identifies the region between the $16^{\text{th}}$ and the $84^{\text{th}}$ percentiles of the probability distribution, with $\log M_p$ being the median.

\section{Results}
\label{sec:results}

\subsection{Simulations}

\begin{figure*}[h!]
    \centering
    \includegraphics[width=0.96\textwidth]{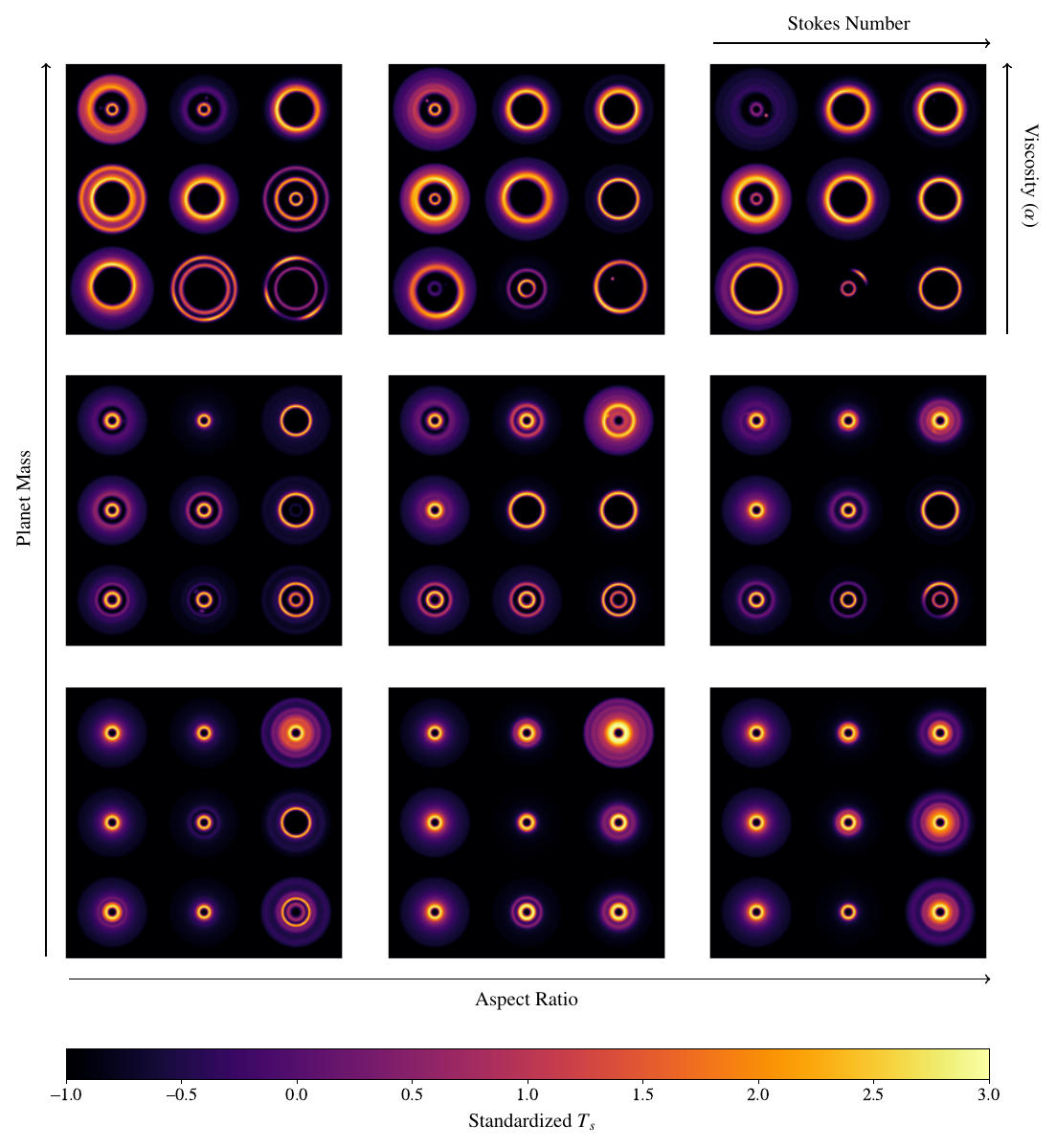}
    \caption{Gallery of some mock observations in our dataset, before the beam convolution, organized to show how the morphology of the substructures varies with the disc’s and planet's physical parameters. The colour bar here refers to the disc brightness temperature, computed with Eq. (\ref{eq:brightT}) and then standardized by subtracting the mean of the image's pixels and dividing by the standard deviation.}
    \label{fig:gallery}
\end{figure*}

It took approximately 15,000 CPU hours to complete all the 1000 simulations.
We collected for each of them the dust density maps at three times: after 500, 1000 and 1500
planetary orbits. These correspond respectively to $1.7 \times 10^5$~yr, $3.5 \times 10^5$~yr and $5.3 \times 10^5$~yr for a planet at 50 au around a solar mass star. We thus post-processed these results as previously explained to obtain a
dataset of $3\times 1000$ synthetic observations. Figure \ref{fig:gallery} shows some of the images collected
in this dataset highlighting the diversity of substructures and their dependence on some of
the disc parameters. It is interesting to observe the presence of some discs without visible
substructures, discs with multiple gaps and many where the planet opened a central hole as well as some with non-axisymmetric structures such as vortices.

\begin{figure}[]
        \includegraphics{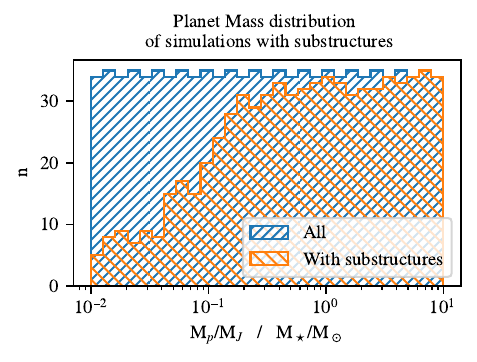}
    \caption{Planet mass distribution in the simulation dataset before (in blue) and after (in orange) removing images without azimuthally symmetric substructures.}
    \label{fig:pmassdist}
\end{figure}

The size of the dataset was reduced to 2196 separate synthetic observations after the removal of those without azimuthally averaged substructures. This operation skews the distribution of planet mass values in the final dataset towards the higher end of the explored mass spectrum (see Fig. \ref{fig:pmassdist}). As we discuss later, the unbalancing of the training dataset might explain the deterioration of the ensemble's performance in the inference of low-mass planets.

\subsection{Training}

The training of each CNN on the five folds took approximately 60 CPU hours for a total cost of 600 CPU hours for the whole ensemble. The low complexity of the architecture that we chose to use did not justify adopting GPUs to train the models as, with the possibility of training all the CNNs in the ensemble in parallel, the time required for this step was sufficiently low to enable a satisfactory exploration of different architectures and hyperparametrizations. However, for considerations regarding the computational costs of our approach, we note that the use of GPUs would have sensitively decreased the training time. Learning curves are shown and discussed in Appendix \ref{app:learncurv}.

\subsection{Results on the test set}

Once the ensemble's training is completed, we tested it on the test set collecting the results of all five folds. 
We remark that the test set contains data homogeneous with those used for training, thus obtained from the same distribution of physical parameters and with the same simulation pipeline.

To evaluate these results, we refer in the next subsections to the standardized error defined as 
\begin{equation}
   \text{Standardized error} = \frac{ \log M_p^\text{target} - \log M_p^\text{predicted}}{\sigma}. \label{eq:stderr}
\end{equation}
We note that while an individual model returns a single physical uncertainty $\sigma_p$, after ensembling, the left $(\sigma_-)$ and right $(\sigma_+)$ uncertainties estimated respectively from the 16$^{\text{th}}$ and 84$^\text{th}$ percentiles of the pdf returned by the ensemble might differ.
Here we call $\sigma$ the mean $\sigma = (\sigma_+ + \sigma_- )/2$. We found $\sigma_+$ and $\sigma_-$ to be very similar in most cases.

\subsubsection{Planet mass estimates}

\begin{figure}[]
      \centering
    \includegraphics{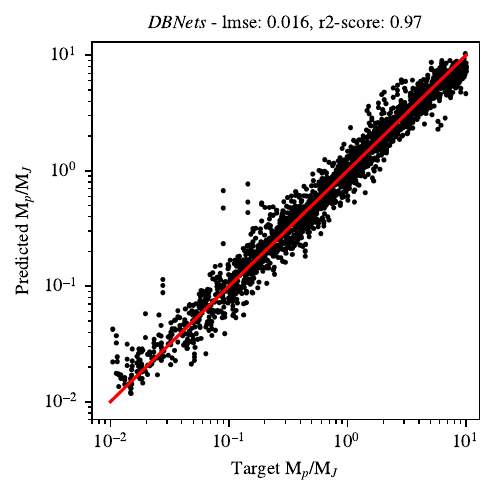}
    \caption{Scatter plot showing the correlation, in the results obtained on the test set, between the target planet mass and the estimate of the ensemble of CNNs (\emph{DBNets}). The red line marks the ideal exact correlation that is targeted.}
    \label{fig:scatter}
\end{figure}

Figure \ref{fig:scatter} shows the masses estimated with our ensemble as a function of the target values highlighting a good correlation. We checked that the results presenting the highest deviation from their target value are those of discs with the highest or lowest aspect ratios.

To quantify the quality of our results we evaluated two uncertainty-agnostic metrics: the lmse, Eq. (\ref{eq:lmse}), and the r2-score.
 The latter is defined as
\begin{equation}
    \text{r2-score} = 1 - \frac{\sum_{i\in\mathcal{D_T}} \left(\log   M_{p,i} - \log M_{\text{targ}, i}\right)^2}{\sum_{i\in\mathcal{D_T}} \left(\log M_{\text{targ}, i} - \left<\log M_{\text{targ}}\right>_{\mathcal{D_T}} \right)^2},
\end{equation}
with $\left< \cdot \right>$ indicating the mean of its argument. The best value that this metric can
assume is 1, while it can, in principle, assume any smaller value. An r2-score equal to 0
corresponds to a model that always predicts the mean value of the targets in the training
set. On the test set, we obtained a lmse of 0.016 and an r2-score of 97\%. Readers can refer to Sect. \ref{sec:linear} for a comparison of these metrics with other methods of planet mass estimation.

The absolute value of the relative error for the planet mass, averaged over the entire test set, is approximately 23\%.

\subsubsection{Estimated uncertainty}
The uncertainties estimated on the test set items vary approximately between 0.08 and 0.4 dex with a median of 0.14 dex.
To evaluate the quality of the uncertainty estimated with the ensemble we computed, for each element of the test set, the previously

\begin{figure}[]
    \centering
\includegraphics{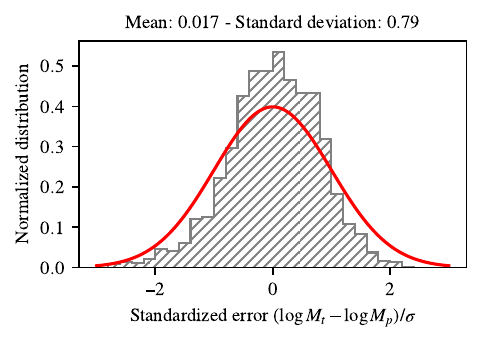}
    \caption{Distribution of the standardized error of \emph{DBNets} mass estimates on the test set. The red line marks the expected Gaussian distribution of mean 0 and variance 1. }
    \label{fig:error}
\end{figure}

\noindent  defined standardized error (Eq. \ref{eq:stderr}).
If we interpret each \emph{DBNets} estimate  $\left\{\log M_p^\text{predicted}, \sigma \right\}$ assuming that the real value of the planet mass is normally distributed around $\log M_p^\text{predicted}$ with variance $\sigma^2$ then the standardized error, Eq. (\ref{eq:stderr}), should be normally distributed with mean 0 and variance 1 for each element of the test set. Since they have to be equally distributed, we can combine all the standardized errors evaluated on the test set to investigate their distribution and benchmark our method of uncertainty estimation.
Figure \ref{fig:error} shows the result of this analysis. The mean of this distribution ($\mu$) is 0.017 in good agreement with the expectation of $\mu=0$ indicating that, averaged across the whole range of planet masses, the ensemble does not systematically overestimate nor underestimate their values. The standard deviation of this distribution $\sigma=0.79$ is, instead, 20\% lower than the expected value $\sigma=1$. This means that the uncertainty returned by the ensemble systematically overestimates the real error. However, this test was performed on the test set that, as anticipated, is homogeneous with the data used during training. 
For this reason, we reasonably expect that the prediction error could increase when analysing different images, such as real observations or a different batch of simulations and we are thus satisfied with the conservative overestimation that we obtained. 
We discuss in Sect. \ref{sec:oodd} if it is sufficient to correctly represent the increased prediction error in applications to out-of-distribution data.

\begin{figure}[]
    \centering
    \includegraphics{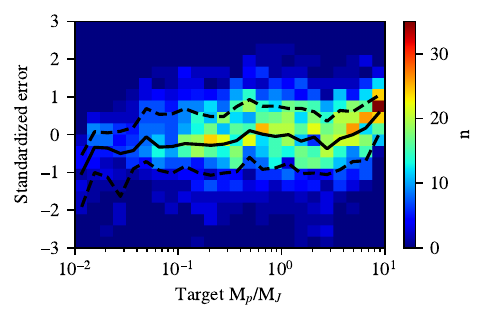}
    \caption{Two-dimensional distribution of the standardized error as a function of the target mass. The black line marks the median of each distribution at a fixed target mass while the dashed lines mark respectively the 16$^\text{th}$ and 84$^\text{th}$ percentiles. The graph shows a distribution in fine agreement with the expected one for planets of masses approximately between $10 \text M_\oplus$ and $6 \text M_J$ (assuming $M_\star = 1 \text M_\odot$). Lower and higher masses are instead, respectively, systematically overestimated or underestimated.}
    \label{fig:errormp}
\end{figure}

\subsubsection{Results dependence on the disc and planet properties}
In Fig. \ref{fig:errormp} we expanded the distribution of the standardized error across the range of planet masses explored in our dataset of simulations. The solid black line represents the median of each distribution of standardized errors at a fixed target planet mass while the dashed lines mark respectively the 16$^\text{th}$ and $84^{\text{th}}$ percentiles. The errors are fairly well distributed around 0 approximately in the range between 10 $M_\oplus$ and 6 $M_J$ (we converted the masses assuming $M_\star = 1 \text{M}_\odot$) while we observe a systematic overestimate and underestimate respectively for lower and higher masses. It is common for deep learning methods to underperform when applied to data at the boundary of the training set domain. However, the magnitude of low-masses overestimations is particularly high, probably as a result of the lower number of synthetic observations with low-mass planets in the training dataset.

\begin{figure*}[p!]
    \centering
    \includegraphics{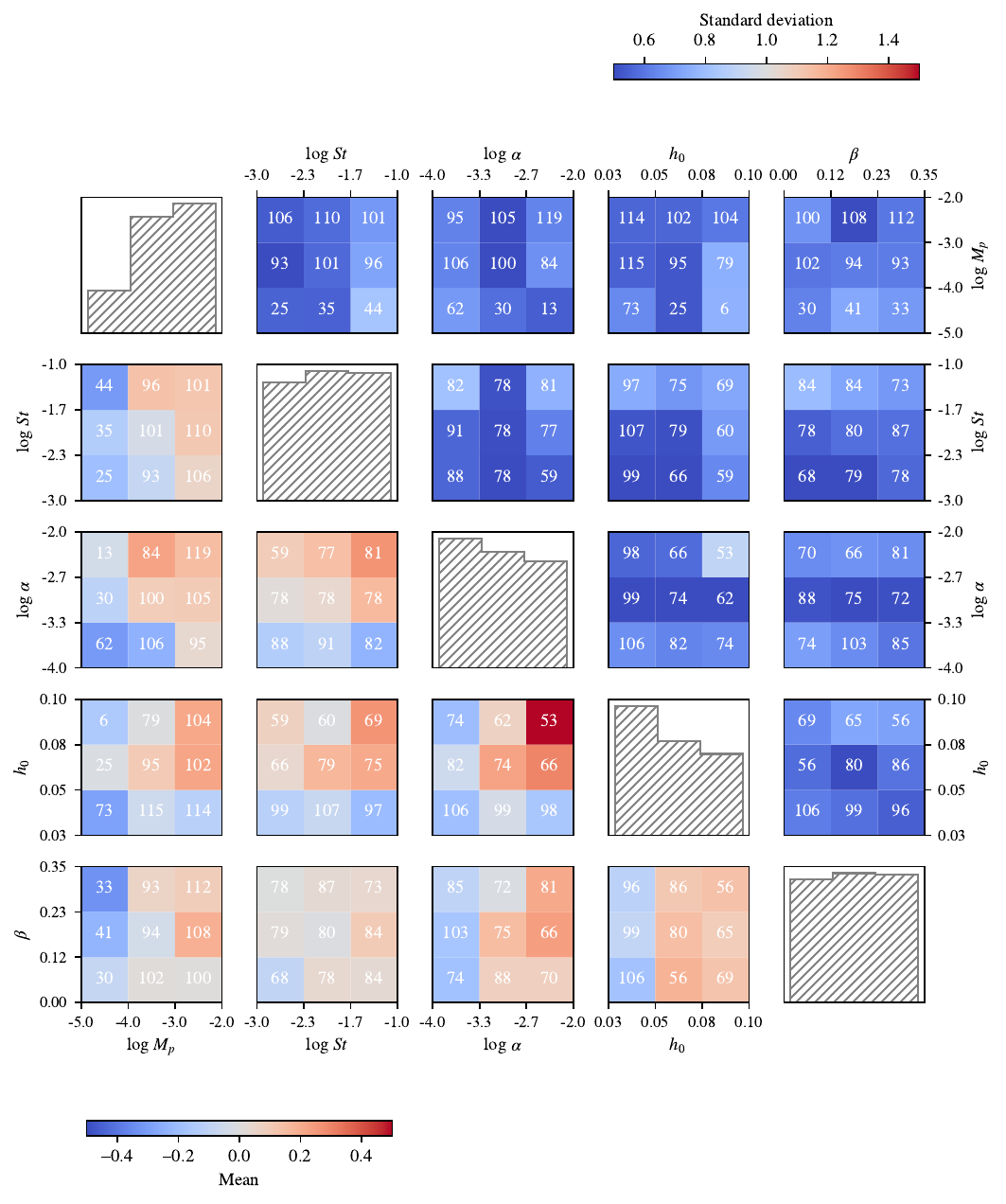}
    \vspace{0.44cm}
    \caption{Mean (bottom left corner) and standard deviation (top right corner) of the distribution of standardized errors evaluated on the test set across the parameter space.
    The histograms on the diagonal are indicative of how the test set images are distributed in the parameter space. We reported inside each cell the number of elements of the test set lying in that region and thus used to compute the moments of the standardized error distribution. } 
    \label{fig:stderrpara}
\end{figure*}

Finally, Fig. \ref{fig:stderrpara} shows how the distribution of the standardized errors varies across the entire parameter space. We note that the standard deviation of its distribution remains systematically lower than 1 across the entire parameter space signalling that it conservatively provides a good estimate of the actual error even when the planet's mass is sistematically over or under-estimated. Observing the mean of this distribution across the parameter space, we instead observe that in some specific regions deviates from the expected null value. This result highlights the existence of several degeneracies in the substructures generated by planets in discs and identifies the affected regions of the parameter space. Apart from the dependence upon the planet mass that we already
discussed, we observe that the disc viscosity and aspect ratio play an important role: at the higher end of both these parameters the planet mass is highly underestimated (positive standardized errors). This is easily explainable observing that in this regime the gap opening is more difficult thus resulting in shallower and narrower gaps that are interpreted by our tool as generated by 
lower-mass planets in `average' (according to our distribution of parameters) discs.

\section{DBNets results on actual observations}
\label{sec:real}

\begin{table*}[h!]
\vspace{0.3cm}
\caption{Catalogue of ALMA observations on which we applied \emph{DBNets}. }
\label{tab:prop}
\resizebox{\linewidth}{!}{%
\begin{tabular}{lrrrrrrrrrrr}
\toprule
\\
\multicolumn{1}{c}{\textbf{\begin{tabular}[c]{@{}c@{}}Disc\\  name\end{tabular}}} & \multicolumn{1}{c}{\textbf{\begin{tabular}[c]{@{}c@{}}Star mass\\  {[}M$_{\sun}$ {]}\end{tabular}}} & \multicolumn{1}{c}{\textbf{\begin{tabular}[c]{@{}c@{}}Distance\\  {[}pc{]}\end{tabular}}} & \multicolumn{1}{c}{\textbf{\begin{tabular}[c]{@{}c@{}}Inclination \\ {[}°{]}\end{tabular}}} & \multicolumn{1}{c}{\textbf{\begin{tabular}[c]{@{}c@{}}Position angle \\ {[}°{]}\end{tabular}}} & \multicolumn{1}{c}{\textbf{\begin{tabular}[c]{@{}c@{}}Resolution\tablefootmark{1} \\ {[}a{]}\end{tabular}}} & \multicolumn{1}{c}{\textbf{\begin{tabular}[c]{@{}c@{}}ALMA\\ band\end{tabular}}} & \multicolumn{1}{c}{\textbf{\begin{tabular}[c]{@{}c@{}}Gap radial \\location(s) a  {[}au{]}\end{tabular}}} & \multicolumn{1}{c}{\textbf{\begin{tabular}[c]{@{}c@{}}Data \\ references\end{tabular}}} & \multicolumn{1}{c}{\textbf{\begin{tabular}[c]{@{}c@{}}Properties \\ references\end{tabular}}} &  \multicolumn{1}{c}{\textbf{\begin{tabular}[c]{@{}c@{}}\cite{Bae2023StructuredDisks} M$_p$ \\ {[}$\text{M}_\text{J}${]}\end{tabular}}}  &  \multicolumn{1}{c}{\textbf{\begin{tabular}[c]{@{}c@{}}\emph{DBNets} M$_p$ \\ {[}$\text{M}_\text{J}${]}\end{tabular}}} \\   \\
\midrule 
AS 209     & $0.83$                   & $121$               & $35$                  & $86$                     & $0.22, 0.02$       & $6$    & $9, 99$                    & a                  & $1$                     & 0.083, 2.05 & $1.20^{+0.50}_{-0.38}$, $^*0.54^{+0.43}_{-0.25}$\\[2mm]
Elias 2-24 & 0.78                   & 136               & 29                  & 46                     & 0.04             & 6    & 57                       & a                  & 1                    & 0.17 & $0.38^{+0.17}_{-0.12}$\\[2mm]
Elias 2-27 & 0.49                   & 140               & 56                  & 117                    & 0.04             & 6    & 69                       & a                  & 1                    &0.06 & $0.40^{+0.23}_{-0.16}$\\[2mm]
GW Lup      & 0.46                   & 155               & 39                  & 38                     & 0.04             & 6    & 74                       & a                  & 1      & 0.03   & $^*0.14^{+0.11}_{-0.06}$            \\[2mm]
HD 142666  & 1.58                   & 148               & 62                  & 162                    & 0.13             & 6    & 16                       & a                  & 1                   &0.3  & $0.59^{+0.28}_{-0.21}$\\[2mm]
HD 143006  & 1.78                   & 165               & 19                  & 169                    & 0.15, 0.06       & 6    & 22, 51                   & a                  & 1                   &19.91, 0.33 & $7.38^{+2.56}_{-1.93}$, $1.16^{+0.71}_{-0.44}$ \\[2mm]
HD 163296  & 2.04                   & 101               & 48                  & 133                    & 0.21, 0.04, 0.02 & 6    & 10, 48, 86               & a                  & 1                    & 0.71, 2.18, 1 & $1.05^{+0.68}_{-0.45}$, $1.92^{+0.77}_{-0.59}$, $^*0.19^{+0.20}_{-0.10}$ \\[2mm]
SR 4       & 0.68                   & 134               & 22                  & 18                     & 0.18             & 6    & 11                       & a                  & 1                   & 2.16 & $1.47^{+0.49}_{-0.37}$ \\[2mm]
DoAr 25    & 0.95                   & 138               & 67                  & 111                    & 0.02             & 6    & 111                      & a                  & 1                 & 0.73   & $0.24^{+0.16}_{-0.10}$  \\[2mm]
Elias 2-20 & 0.48                   & 138               & 49                  & 153                    & 0.08             & 6    & 25                       & a                  & 1                     & 0.57 & $0.22^{+0.11}_{-0.08}$\\[2mm]
RU Lup     & 0.63                   & 154               & 19                  & 121                    & 0.06             & 6    & 29                       & a                  & 1                     & 1.18 & $0.29^{+0.16}_{-0.10}$\\[2mm]
IM Lup     & 0.89                   & 158               & 48                  & 143                    & 0.03             & 6    & 117                      & a                  & 1                    &0.039 & $^*0.21^{+0.16}_{-0.09}$ \\[2mm]
Sz 114    & 0.17                   & 162               & 21                  & 165                    & 0.19             & 6    & 24                       & a                  & 1                   &0.02 & $^*0.06^{+0.04}_{-0.03}$ \\[2mm]
Sz 129     & 0.83                   & 161               & 34                  & 151                    & 0.07             & 6    & 41                       & a                  & 1                     & 0.03 & $0.16^{+0.09}_{-0.06}$\\[2mm]
HL Tau     & 1                      & 140               & 47                  & 138                    & 0.14, 0.05, 0.03 & 7    & 13.1, 33, 68.6           & b                  & 2                   &  0.35, 0.17, 0.26 & $1.01^{+0.42}_{-0.32}$, $0.82^{+0.43}_{-0.31}$, $0.55^{+0.31}_{-0.20}$\\[2mm]
HD 169142  & 1.65                   & 117               & 5                   & 13                     & 0.08, 0.03, 0.02 & 6    & 17, 50, 64               & d                  & 4,5,6,7,8             & 2.4, 2.4, 0.03 & $^*7.57^{+5.59}_{-3.82}$, $0.68^{+0.38}_{-0.25}$, $^*0.96^{+0.85}_{-0.51}$\\[2mm]
HD 97048   & 2.5                    & 183               & 40                  & 3                      & 0.07             & 7    & 130                      & c & 7,8                   & 2 & $0.82^{+0.38}_{-0.26}$\\[2mm]
Lk Ca 15   & 1.25                   & 159               & 50                  & 62                     & 0.11, 0.06       & 6    & 36, 70                   & e                  & 3, 9                  & 0.47, 0.02 & $3.45^{+1.13}_{-0.86}$, $0.71^{+0.33}_{-0.27}$\\[2mm]
TW Hya     & 0.8                    & 56                & 56                  & -8                     & 0.03, 0.01       & 7    & 20, 81                   & f                  & 3                     & 0.15, 0.08 & $0.80^{+0.39}_{-0.26}$, $^*2.77^{+5.89}_{-1.80}$\\[2mm]
CI Tau     & 0.9                    & 159               & 50                  & 11                     & 0.24, 0.08, 0.03 & 6    & 14, 43, 119              & g & 10                  & 0.75, 0.15, 0.4 & $3.10^{+1.14}_{-0.79}$, $0.70^{+0.32}_{-0.25}$, $0.56^{+0.31}_{-0.21}$ \\[2mm]
CR Cha      & 1.5                    & 187               & 31                  & 36                     & 0.08             & 6    & 90                       & h & 11                    & 0.31 & $^*0.58^{+0.46}_{-0.27}$\\[2mm]
DS Tau     & 0.83                   & 159               & 65                  & -19                    & 0.29             & 6    & 33                       & i  & 12                  & 3.5 & $1.28^{+0.51}_{-0.43}$\\[2mm]
DM Tau     & 0.5                    & 145               & 33                  & -25                    & 0.03             & 3    & 70                       & l & 13              &0.333 & $0.26^{+0.15}_{-0.10}$\\[2mm]
DL Tau & 0.98 & 159 & 45 & 52 & 0.24, 0.14, 0.11 & 6 & 39,67,89 & i & 14 & 0.11, 0.08, 0.33 & $0.17^{+0.10}_{-0.07}$, $1.23^{+0.67}_{-0.49}$, $0.55^{+0.32}_{-0.22}$\\[2mm]
DN Tau & 0.52 & 128 & 35 & 79 & 0.19 & 6 & 49 & i & 14 & 0.009 & $^*0.20^{+0.15}_{-0.09}$\\[2mm]
FT Tau & 0.34 & 127 & 35 & 122 & 0.28 & 6 & 25 & i& 14 & 0.15 & $0.19^{+0.11}_{-0.07}$\\[2mm]
GM Aur & 1.32 & 159 & 53 & 57 & 0.05 & 6 & 67 & m & 15 & 0.1 & $0.29^{+0.13}_{-0.09}$\\[2mm]
GO Tau & 0.36 & 144 & 54 & 21 & 0.15, 0.10 & 6 & 59,87 & i & 14 & 0.057, 0.07 & $0.22^{+0.14}_{-0.08}$, $^*0.12^{+0.13}_{-0.06}$\\[2mm]
HD 107146\tablefootmark{2} & 1 & 27.5 & 19 & 153 & 0.12 & 6 & 80 & n & 16 & 0.1 & $0.76^{+0.50}_{-0.31}$\\[2mm]
IQ Tau & 0.5 & 131 & 62 & 42 & 0.22 & 6 & 41 & i & 14 & 0.065 & $0.11^{+0.06}_{-0.05}$\\[2mm]
MWC 480 & 1.91 & 161 & 36.5 & 147.5 &  0.16 & 6 & 73 & i & 14 & 1.3 & $1.06^{+0.45}_{-0.36}$\\[2mm]
MWC 758 & 1.5 & 160 & 21 & 62 & 0.10 &7 & 30 & o & 17 & 1.5 & $7.85^{+2.88}_{-2.19}$\\[2mm]
UZ Tau & 0.39 & 131 & 56 &90  & 0.10  & 6 & 69 &  i & 14 & 0.023 & $0.15^{+0.10}_{-0.06}$\\[2mm]
\bottomrule
\end{tabular}
} 
\tablefoot{\textit{Data references:} (a) \href{https://almascience.eso.org/almadata/lp/DSHARP/}{DSHARP Data Release}, (b) \href{https://almascience.eso.org/alma-data/science-verification}{ALMA Science Verification Data}, (c) \cite{Pinte2019HD13CO}, (d) \cite{Perez2019DustRing}, (e) \cite{Facchini2020AnnularJ1610}, (f) \cite{Andrews2016RINGEDDISK}, (g) \cite{Clarke2018High-resolutionAu}, (h) \cite{Kim2020TheDisk}, (i) \cite{Long2018GapsRegion}, (l) \cite{Hashimoto2021ALMATau}, (m) \cite{Huang2020ADisk}, (n) \cite{Marino2021ConstrainingEdges}, (o) \cite{Baruteau2019DustPlanets}, (p) \cite{zagariaprep2}.
\textit{Properties references:} (1) \cite{Zhang2018TheInterpretation}, (2) \cite{Lodato2019TheDiscs}, (3) \cite{Dong2017WhatPlanet}, (4) \cite{Toci2019Long-lived169142}, (5) \cite{Perez2019DustRing}, (6) \cite{Gratton2019Blobs169142}, (7) \cite{Pinte2019KinematicDisk}, (8) \href{https://exoplanets.nasa.gov/}{NASA exoplanets catalogue}, (9) \cite{Facchini2020AnnularJ1610}, (10) \cite{Clarke2018High-resolutionAu}, (11) \cite{Kim2020TheDisk}, (12) \cite{Veronesi2020IsPlanet}, (13) \cite{Wang2021ArchitectureGap}, (14) \cite{Long2018GapsRegion}, (15) \cite{Huang2020Large-scaleLup}, (16) \cite{Marino2021ConstrainingEdges}, (17) \cite{Dong2018MultipleSystems}.}
\tablefoottext{*}{Unreliable \emph{DBNets} mass estimates.}\\
\tablefoottext{1}{ This column is aimed at allowing a comparison with the resolution of the synthetic observations (0.12a). We report the major standard~deviations of the two-dimensional Gaussians that approximate the observational beams. Values are expressed in units of the gap radial location $a$.}\\
\tablefoottext{2}{ Note that, unlike the others, this is a debris disc.} 
\vspace{0.4cm}
\end{table*}

As a final step of our work, we decided to apply the tool that we developed to a conspicuous group of actual protoplanetary disc observations. In this section we discuss the data used and the results obtained.

\subsection{Data selection}
We referred to \cite{Bae2023StructuredDisks} to identify suitable protoplanetary discs candidates meeting two requirements: availability of high-resolution observation of the dust thermal emission in ALMA Band 6 or 7 and the presence of observable axisymmetric substructures in these data. 

To properly apply \emph{DBNets}, a few additional properties of the object under consideration are required. First of all, it is necessary to know the disc inclination and position angle to properly deproject the observation into a face-on image. For the same reason, a guess on the planet's location is needed either in physical units with an estimate of the disc distance or as the angular separation with respect to the disc centre. With this information, the input image has to be rescaled to match the pixel scale of the training data in units of the planetary orbital radius.
 We provide in our tool a preprocessing module that takes care of these steps given the necessary geometrical information. Additionally, because we observed that our tool works best when applied on observations
with the same resolution of our training dataset (i.e. 0.12$r_0$, see Appendix \ref{app:ood}), we reduced to this value the resolution of all the better-resolved observations convolving them with an appropriate Gaussian beam. 
Lastly, the star mass is required to convert the estimated planet-star mass ratio to a planet mass in physical units.

Table \ref{tab:prop} shows the objects that we considered in this study with the relative properties and references. When multiple gaps are present in a disc, we applied our tool once per gap conjecturing each time the presence of a planet in the middle of the gap under consideration.

\subsection{Results}

In Fig. \ref{fig:preds} we reported the results obtained in ascending order (from top to bottom) with respect to the uncertainty of \emph{DBNets} results. The black points mark the median of the estimated pdf for the $\log M_p$ while the error bars delineate the confidence interval between the 16$^\text{th}$ and 84$^\text{th}$ percentiles. We reported in the same figure some proposed planet masses available in literature for the putative planets in these substructures. 
The ring and square markers indicate estimates obtained from the analysis of the
observed substructures through empirical relations or hydrodynamical simulations while
the stars mark estimates obtained from kinematical data. We remark that these values
from other works are not the true target planet masses but just guesses obtained with
different methods and possibly different models and assumptions on the disc properties.
Indeed, note that in a few cases when we show values from different works, these
often do not agree. See for example AS 209, Elias 2-24, or Elias 2-20. 
The total CPU time taken by \emph{DBNets} to estimate these masses is $\sim 7$ seconds, demonstrating the efficiency of deep learning methods in accelerating the analysis of astrophysical data.

\begin{figure*}[]
    \centering
      \hspace{0.5cm}
      
    \includegraphics{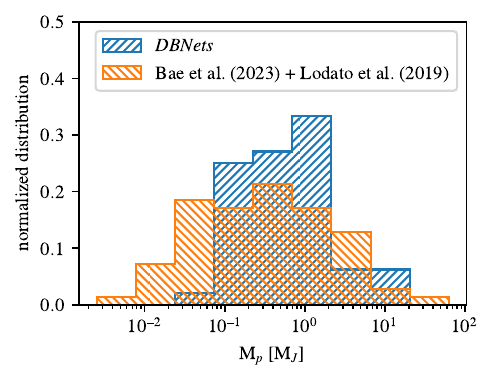}
    \hspace{0.2cm}
    \includegraphics[]{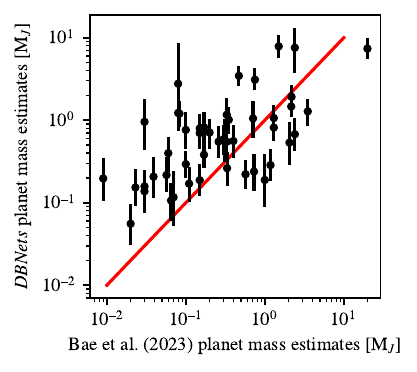}
    \caption{Comparison of \emph{DBNets} and literature \citep{Bae2023StructuredDisks, Lodato2019TheDiscs} planet mass estimates for the sample of observations outlined in Table \ref{tab:prop}. The left panel compares the normalized distributions of the inferred planet masses. The right panel shows explicitly, for each putative planet, the differences between \emph{DBNets} mass estimates and those reported in \cite{Bae2023StructuredDisks}. The red line marks the identity relation for reference.}
    \label{fig:dist}
\end{figure*}

In support of our
results, we observe that in most occasions the confidence intervals that we obtained for
the planets’ masses include at least one of the values proposed in separate works and generally follow their trend. See also the left panel of Fig. \ref{fig:dist} for a comparison of the mass distributions and the right panel for a more direct view of the differences between \emph{DBNets} planet mass estimates and those reported in \cite{Bae2023StructuredDisks}. The few instances where our findings strongly diverge from the
literature typically involve cases where the planetary mass values reported from external
studies (mainly from \citealt{Lodato2019TheDiscs} whose estimates are only reported in Fig. \ref{fig:preds}) fall within the low end of the mass spectrum. This is consistent with the fact that, in all the tests that
we performed, we found the estimation error of \emph{DBNets} increasing for low-mass planets. Our tool can be integrated and compared with
available literature and different planet characterization techniques to asses if a low-mass
planet in a given disc is more likely and decide accordingly on the significance of the
\emph{DBNets} measure. Also, once one has the \emph{DBNets} estimate, one can do targeted simulations for a specific system.
The red dashed line separates the mass estimations according to the uncertainty threshold of 0.25 dex that we subsequently discuss in Sect. \ref{sec:oodd}. We anticipate, for the purpose of this section, that estimations below this line must be treated with additional caution as the above-threshold uncertainties may signal that the input data is too different from the information used for training the deep learning models. We were not able to find a specific physical feature of these disc images, such as the resolution or disc inclination, that could be confidently tied to the higher uncertainty. However, we observed that in most of these cases, the observation is either noisy or with very shallow substructures. Mass estimates inferred from dust images whose resolution is significantly worse than that simulated in our training dataset ($\gtrsim 0.2r_0$) do not present systematic differences in their uncertainty or in their deviation from the literature (see Appendix \ref{app:resreal}).  The same is also true for highly inclined discs ($\gtrsim 60$°).

A more quantitative comparison can be carried out analysing the distribution of the deviations between \emph{DBNets} and literature \citep{Bae2023StructuredDisks,Lodato2019TheDiscs} estimates, in units of \mbox{\emph{DBNets}} uncertainties defined as the standardized error, Eq.~(\ref{eq:stderr}), with literature values as targets. Over the considered sample, this quantity has mean $-1.4$ and root mean square $3.1$. The left panel of Fig. \ref{fig:dist} shows the 2 distributions of \emph{DBNets} and literature \citep{Bae2023StructuredDisks, Lodato2019TheDiscs} mass estimates. We observe, as already found computing the mean deviation, that \emph{DBNets} returns typically higher masses. We also observe that the 
distribution of \emph{DBNets} mass estimates is narrower with a standard deviation (computed with the log M$_p$ values) $\sim 35\%$ lower than literature values. We checked that this cannot be attributed to our bias against low-mass planets, which are more prone to overestimations of their masses.
We performed a Kolmogorov-Smirnov test to compare the two samples rejecting the hypothesis that they are drawn from the same underlying distribution (p-value: $4\times10^{-4}$).

\subsection{Implications of our results}

\begin{figure*}[]
    \centering
\includegraphics[width=0.9\textwidth]{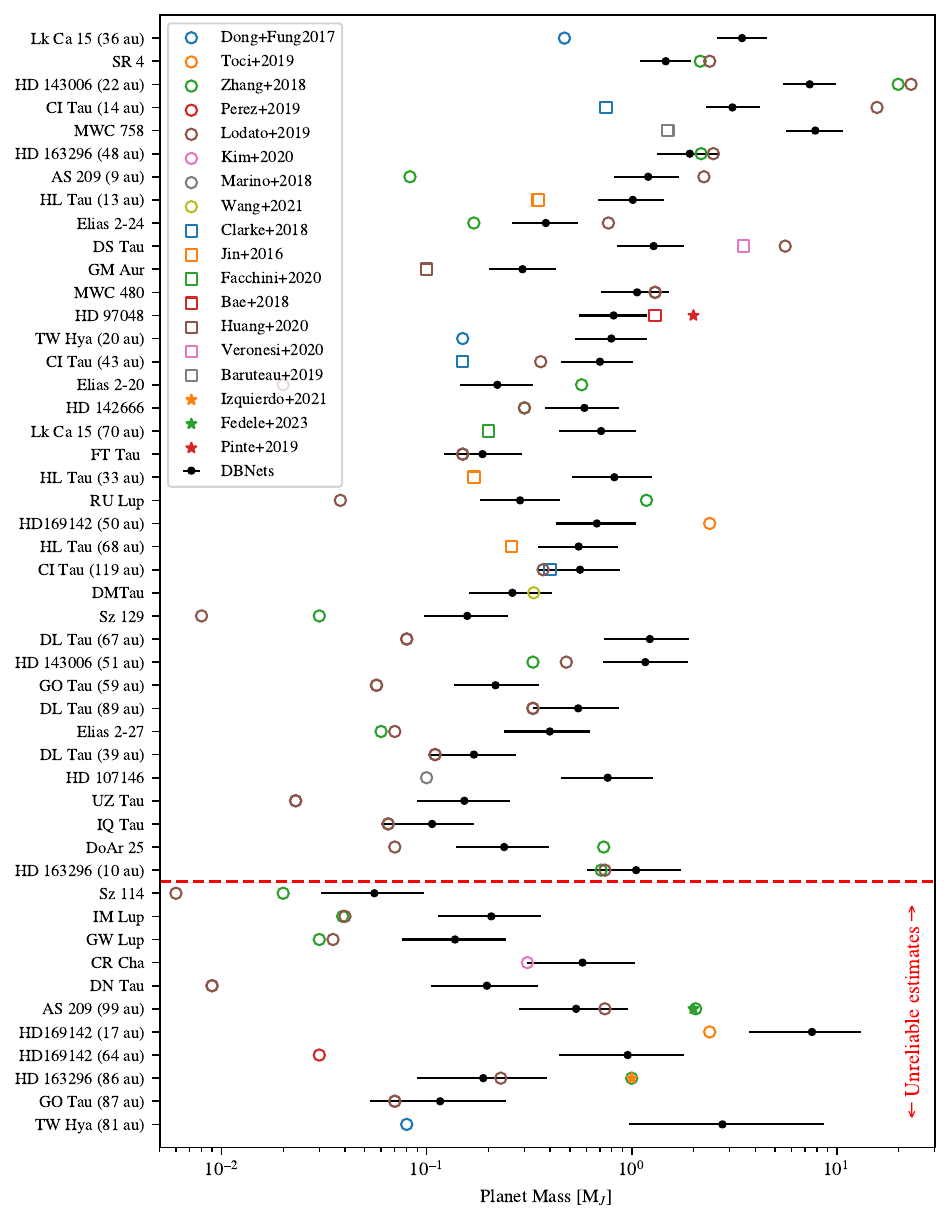}
    \caption{Summary of \emph{DBNets} results on the collected catalogue of observations and a comparison with proposed planet masses available in the literature. The rings and squares indicate estimates obtained examining dust substructures while the stars refer to kinematical detections. The results are presented (from top to bottom) in ascending order with respect to the uncertainty of the \emph{DBNets} results. The red dashed line separates results according to the uncertainty threshold of 0.25 dex. References: \cite{ Dong2017WhatPlanet, Toci2019Long-lived169142, Zhang2018TheInterpretation, Perez2019DustRing,  Lodato2019TheDiscs}; \cite{ Kim2020TheDisk,  Marino2018AALMA, Wang2021ArchitectureGap, Clarke2018High-resolutionAu,   Jin2016MODELINGINTERACTIONS, Facchini2020AnnularJ1610, Bae2018DiversePlanet, Huang2020ADisk,  Veronesi2020IsPlanet, Baruteau2019DustPlanets,  Izquierdo2022ADiscminer, Fedele2018ALMA209, Pinte2019KinematicDisk}.}
    \label{fig:preds}
\end{figure*}

The 3 kinematics mass estimates, see Fig. \ref{fig:preds}, are all significantly higher than \emph{DBNets} estimates with a mean deviation of $2.25$ in units of \emph{DBNets} uncertainties. This deviation is larger and in the opposite direction with respect to the discrepancy of \emph{DBNets} measures with literature estimates inferred from dust substructures. Although our sample is small, our results are in agreement with the already observed \citep{Pinte2020NineGaps} tendency of planet masses inferred from kinematical signatures to be higher than those obtained modelling disc substructures. 

\begin{figure*}[]
\centering
    \includegraphics{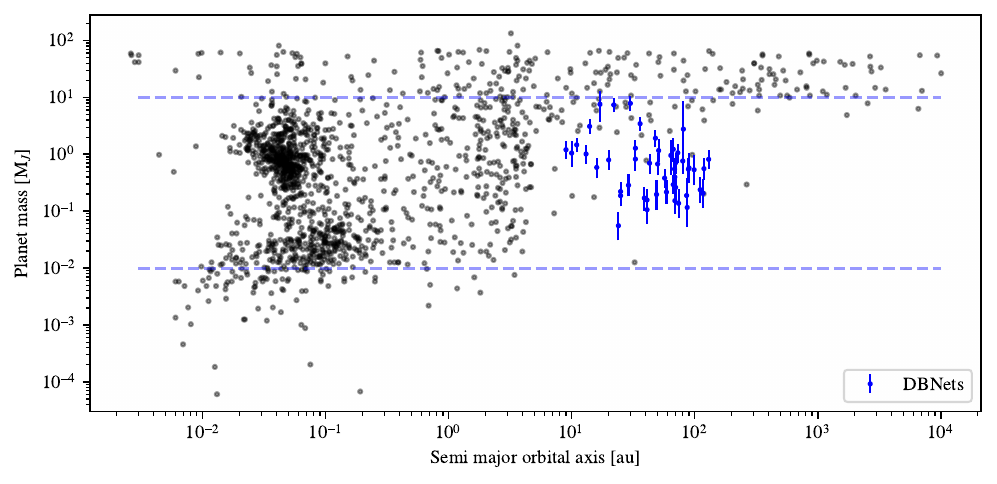}
    \caption{Mass and semi-major orbital axis of the over
5000 exoplanets’ confirmed detections (grey points, data from \href{http://exoplanet.eu}{exoplanet.eu}). The blue error bars are proposed planets in protoplanetary discs characterized
with \emph{DBNets}. The blue dashed lines mark the range of planetary masses spanned in the dataset used to train our tool.}
    \label{fig:madbnets}
\end{figure*}
\noindent
In Fig. \ref{fig:madbnets}, we reported in the
$M_\text p - a$ plane all the exoplanets we conjectured in these observations and characterized
with \emph{DBNets}. This plot showcases the ultimate purpose of our
tool remarking on the importance of revealing planets in protoplanetary discs as a way to
explore regions of the parameter space that cannot be explored with traditional techniques
for the detection of exoplanets. For instance, the characterization of young Jupiter-mass planets at 10-100 au from the host star may help tackle the question of the origin of hot Jupiters providing a way to test and constrain migration scenarios \citep{Lin1996OrbitalLocation, Dawson2018OriginsJupiters}. Assuming that each gap of the considered sample hosts a companion, we characterized here a population of planets at 10 to 130 au from their host star with masses between $10^{-2}$ and 10 M$_\text J$. Although higher masses are easier to observe at these star distances, we found that most (68\%) of the characterized planets are sub-Jupiters thus indicating that super-Jupiter planets cannot be a particularly significant population. This was already suggested by some surveys and observations with non-detections that put upper limits on the masses of the hypothesized \mbox{planets (\citealt{Reggiani2016TheOrbits}; }\citealt{Nielsen2019TheAu, Vigan2021TheSHINE, Asensio-Torres2021Perturbers:Disks}). 
Our results provide further and stronger evidence for this statement presenting the most extensive set of mass 
estimates obtained from a common method that relies on the same assumptions and models. Additionally, we remark that our tool is able to account for degenerate substructures within the considered parameter space and, in this regard,
 we highlight the widely extended ranges of parameters' values explored (see Table \ref{tab:par_table}). 

\label{sec:stability}
Some recent works have raised questions regarding the stability of planetary systems conjectured from protoplanetary discs' dust substructures \citep{Tzouvanou2023DoPlanets}. We checked that most of the systems we proposed in this work are stable according to \cite{Gladman1993DynamicsPlanets} and \cite{Chambers1996TheSystems} criteria (see Appendix \ref{app:stability}). Note, however, that we never selected more than three gaps in the same disc.

\section{Discussion}
\label{sec:discussion}

\subsection{Comparison with a simpler method}
\label{sec:linear}
Inspired by the empirical criteria used by \cite{Rosotti2016TheObservations}, \cite{Kanagawa2016MassWidth}, \cite{Lodato2019TheDiscs} and \cite{Zhang2018TheInterpretation} we manually measured the width of the main gap (closer to the planet position) in the dust for each simulation and then derived an empirical formula to link this gap feature with the
mass of the gap-opening planet. We measured the gap width on the azimuthally averaged profile of the dust surface density selecting, as the inner and outer boundaries, the radii where the density profile reaches half the value between the minimum inside the gap and, respectively, the inner and outer maxima.  To infer our empirical formula, we fitted a power law between the gap width, the planet mass, the Stokes number, the disc aspect ratio and its viscosity obtaining the following:
\begin{equation}
    \frac{M_p}{M_\star} = 0.003  \left( \frac{\Delta}{r_0} \right)^{3.6}  \left( \frac{\alpha}{10^{-3}}\right)^{0.05} \left( \frac{h_0}{0.05}\right)^{-0.14}  \left( \frac{St}{10^{-2}}\right)^{-0.14}. \label{eq:empirical}
\end{equation}

We observe that the strongest dependence that we found with the planet mass is on the gap width $\Delta$ while the dependence on the other parameters is very shallow.

\begin{figure}[t]
    \centering
    \includegraphics{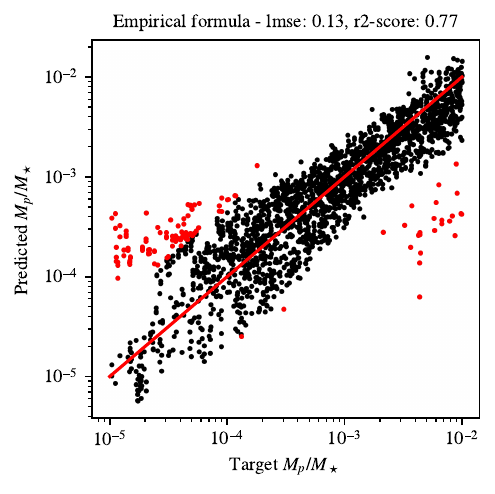}
    \caption{Results of the application on the entire dataset of the empirical formula Eq. (\ref{eq:empirical}). The scatter plot shows the correlation between the target planet mass and that estimated with Eq.  (\ref{eq:empirical}). The red line marks the ideal exact correlation that is targeted.  The evident outliers characterized by an error on the logarithm of the predicted planet mass greater than 0.7 dex are highlighted in red.  }
    \label{fig:completereg}
\end{figure}

The scatter plot
in Fig. \ref{fig:completereg} shows the correlation between the target planet masses of our dataset and
those predicted with this empirical relation. Note the higher scatter with respect to the same plot  with \emph{DBNets}' results (Fig. \ref{fig:scatter}). In this case, too, we computed both the lmse
and the r2-score obtaining respectively 0.13 and 77\%.
We can adopt the square root of the lmse, 0.36 dex, as the uncertainty of this empirical formula. This value corresponds to the $99.7^\text{th}$ percentile of the distribution of \emph{DBNets} uncertainties on the test set. 

The data points at both ends of the mass range with significant over and underestimated predicted values, highlighted in red in Fig. \ref{fig:completereg}, cannot be linked to any specific region of the parameter space. However, we checked that some of these points correspond to simulations that develop complex substructures with several maxima and minima in the azimuthally averaged radial density profile. In these cases, our definition of gap width fails to capture the actual width of the substructure that correlates with the planet mass. To account for this problem in our analysis we computed our evaluation metrics excluding all the outliers with an error on the logarithm of the predicted planet mass greater than 0.7 dex (red points in Fig.~\ref{fig:completereg}). This results in a lmse of 0.08 and an r2-score of 85\%.

Comparing these plots and the metrics' scores reported in Table \ref{tab:metrics} we can confidently affirm that our tool outperforms empirical relations in the estimation of planets' masses from the observed morphologies of tidally induced substructures in protoplanetary discs. To establish a reference point, we computed and reported in the same table scores for the considered metrics in the case of a) random guessing the planet mass with a random value in the dataset’s planet mass range or b) returning for its estimate always the mean value of the range of values explored in the dataset of synthetic observations. The difficulties encountered in defining a universally measurable gap property able to constrain the diversity of planet-induced substructures further strengthen our argument in favour of deep learning methods.

\begin{table}
    \centering
    \vspace{0.2cm}
    \caption{Comparison of DBNets performance with other mass inference methods through the two metrics lmse and r2-score. }
    \label{tab:metrics}
    \begin{tabular}{l|cc}
    \toprule
       Method  & lmse & r2-score \\
       \midrule
       Random guess   & 12.3 & -22 \\
       Mean guess & 0.53 & 0\\
       Empirical formula (entire dataset)& 0.13 & 0.77 \\
       Empirical formula (without outliers) & 0.08 & 0.85 \\
       DBNets & 0.016 & 0.97 \\
       \bottomrule
    \end{tabular}
    \tablefoot{
        Empirical formulae refer to the empirical relation, Eq. (\ref{eq:empirical}), between the gap width and the disc and planet parameters. In the other two methods, the planet mass is predicted with a random value in the dataset's planet mass range (random guess) and with the mean of the target masses (mean guess). These methods serve as references to help interpret the metrics' values.
    }
\end{table}

\subsection{Sources of DBNets quantified uncertainties}
\label{sec:modelvsphy}
We introduced in Sects. \ref{sec:cnnarch} and \ref{sec:ensemble} the concepts of physics and model uncertainty and outlined how we designed our pipeline to provide an estimate of both these contributions.
To be able to evaluate the two  separately, based on our concepts of physics and model uncertainty, we define, for each input to our tool, the physics uncertainty as the
mean of the $\sigma_{p,i}$ returned by the CNNs in the ensemble and the model uncertainty as the
standard deviation of all the $\log   M_{p,i}$ values.
With these definitions, we computed on the test set both the model and physics uncertainties for each synthetic image comparing their values.

\begin{figure*}
    \centering
    \includegraphics{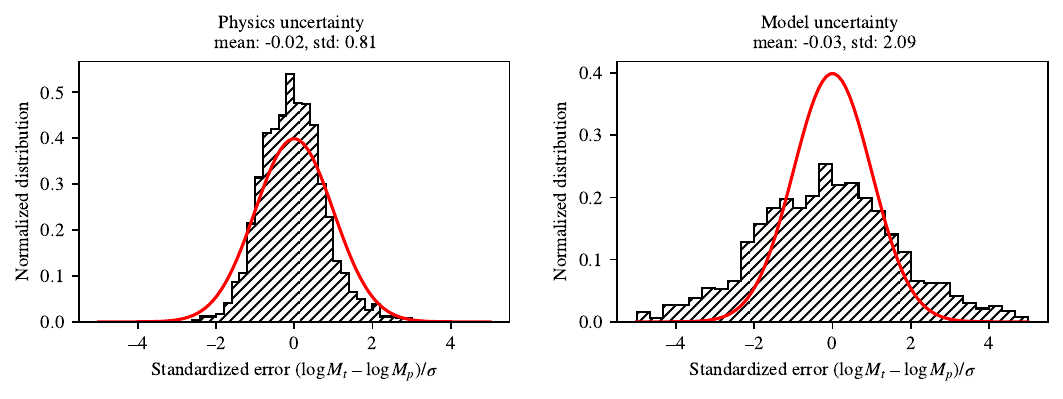}
    \caption{Comparison between the distributions of the standardized errors of \emph{DBNets} mass estimates on the test set computed assuming, as the uncertainty, the {physics} (left panel) and {model} (right panel) contributions defined in Sect. \ref{sec:modelvsphy},  respectively. The red lines mark the expected Gaussian distributions of
mean 0 and variance 1. }
    \label{fig:pvsmodstderr}
\end{figure*}

We found that the uncertainties obtained from the single CNNs dominate over
the model uncertainty due to the small difference between the estimates of the single individuals in the ensemble. Additionally, we observed that the two are usually slightly correlated and that the model uncertainties have a higher scatter varying approximately within two orders of magnitudes instead of one as the physics uncertainty.
Figure \ref{fig:pvsmodstderr} shows the
standardized errors of the predictions on the test set computed using these uncertainties separately.
It is clear from these plots that the model uncertainty alone is not sufficient to estimate
the prediction error adequately. This result highlights the importance, in the use
of the ensembling method, of constructing CNNs that are singularly able to provide an
uncertainty estimate instead of relying solely on the variability of the predictions among
the ensemble’s individuals.

\subsection{Out-of-distribution data detection}
\label{sec:oodd}

The term out-of-distribution (OOD) refers to data drawn from a different distribution with respect to those used for training the deep learning models. For example, data collected at a different time or obtained from different models and assumptions. Deep learning methods, in fact, strongly rely on the data exposed to them during training and improving the generalization capabilities of these tools is challenging. 
In this work, we used highly simplified hydrodynamic and radiative transfer models to obtain the synthetic observations of our training set. The actual data, on which we aim to apply our tool, are characterized by several additional features and non-idealities that were not represented in the training dataset. We developed and trained the deep learning models at the basis of our tool adopting solutions with the aim of being able to generalize the results moving from simulations to actual data reducing as much as possible the generalization error. Completely eliminating this issue, however, is not possible. Indeed, we were only partially able to mitigate the loss of performance of \emph{DBNets} when it is applied to OOD data. For this reason, it is useful to establish an OOD detection method that can be used to validate, and thus accept or reject, a planet mass estimate provided by our tool based on the detection of input data which are considered by the CNNs too different from the data used for their training.

For this purpose, we established an acceptance/rejection threshold of the estimated uncertainty at 0.25 dex which proved effective in detecting OOD input data. To select this threshold we tested our tool on several groups of data completely or partially inhomogeneous with the training dataset. To start, we generated
1000 noise images from a standard Gaussian distribution. We assessed that when applied to this noise our tool systematically returns a significative higher uncertainty than when applied to the test set, in this case well above the 0.25 dex threshold with a median of 0.68 dex. We observed that this increase is due to both the physics and model contributions that we discussed in Sects.~\ref{sec:cnnarch} and \ref{sec:ensemble}.
Secondly, we tested \emph{DBNets} on 1000 synthetic observations of smooth discs without substructures observing that, although decreasing (median of 0.28 dex), the uncertainties that we obtained were still always above the 0.25 dex threshold.
Finally, we studied the effect of several features and non-idealities that might affect actual observations and that we did not include in our sample of synthetic data. For each one of these we applied the following procedure: 1) we modified our test set including in the synthetic observations the feature under consideration; 2) we applied the tool to these modified data; 3) knowing the target values, we performed the same evaluations that we carried out on the original test set analysing and comparing the results. Specifically, we looked at how the distribution of standardized errors changes as the considered non-ideality becomes more relevant. As discussed in Sect. \ref{sec:results}, a deviation of the mean of this distribution from 0 or an increase of its standard deviation above 1 means that the masses and uncertainties provided by DBNets are not, on average, good estimates. We assessed in this way the scope of applicability of our tool and checked if the uncertainty threshold is effective in detecting each one of these deviations from the training dataset. Indeed, we found that, for most of the cases that we studied, when DBNets estimates become unreliable the uncertainty is usually higher than the threshold of 0.25~dex that we selected, motivating our choice of this rejection criteria. Note, however, that these results are true on average but they might not be for specific cases. For example, estimates from marked substructures, usually due to more massive planets, may be less affected by noise or deprojection effects. We can observe this in our results on actual observations: we found (as reported in Appendix \ref{app:ood}) that for discs more inclined than 60° \emph{DBNets} is, on average, less reliable and that the uncertainty estimated is usually higher than 0.25 dex, but, when we applied \emph{DBNets} to the actual observations of the four discs in our sample more inclined than 60°, we always obtained uncertainties lower than 0.25 dex. However, the two systems, IQ Tau and DoAr 25, for which we estimate the lower planet-star mass ratio have uncertainties close to the 0.25 dex threshold while we estimate higher mass ratios and lower uncertainties for DS Tau and HD 142666. 

We discuss here our specific findings on the most relevant sources of error that we evaluated. Additional studied effects are reported in Appendix \ref{app:ood}.

{\bfseries Optical depth}: we found that our tool is reliable, with results similar to those obtained on the original test set, as long as the disc remains optically thin. We observed a significant worsening of the tool accuracy, especially for low-mass planets, when the discs start transitioning towards the optically thick regime. In this case, we found that the uncertainty estimated with \emph{DBNets} does not increase above the threshold of 0.25 dex and thus this deviation from the conditions of training data cannot be easily detected.

{\bfseries Observational noise}: we evaluated how the addition of white Gaussian noise to our synthetic observations affects the predictions of our tool. We found that the results are not affected as long as the signal-to-noise ratio is greater than approximately 10. The estimated uncertainty increases significantly above the 0.25 dex threshold when the noise starts affecting our tool performance.

{\bfseries Evolution}: we evaluated the effect of the evolution, over time, of the disc substructures due to the ongoing planet-disc interaction. To do that, we first trained our models only on synthetic observations taken, for each hydrodynamical simulation, from the same time step. We thus tested these models on the snapshots taken at different times and compared the metrics and qualitative performance. We found that the substructures evolve significantly over time leading, for example, to planet mass underestimate when the models trained on the snapshots after 1500 orbits are applied to those generated from the results after 500 orbits. Following these tests, we observed that this degeneracy could be partially resolved by training our tool on snapshots from different times. This did not hinder the convergence of the training procedure but allowed us to obtain better performance on all the synthetic observations in the test set. Hence, all the results presented in this paper are based on the tool trained on snapshots taken at different times.

\subsection{Caveats}
\label{sec:caveats}

\subsubsection{Multiple gaps}
\label{sec:multip}
In the dataset of hydrodynamical simulations that we used in this work to train the deep learning models, we only placed one planet in each disc. In actual discs, the presence of multiple planets is also likely. In the simplest scenario, each planet of a multiple system may be thought responsible for the independent formation of one localized gap at its location with the same morphology that it would produce in a disc without other companions. When we applied our tool to discs with multiple gaps in Sect. \ref{sec:real}, this is the assumption we made. However, the presence of multiple planets within the same disc can affect the formation and morphology of all the disc substructures. Furthermore, multiple gaps can also be produced by a single planet \citep{Dong2018MultipleSystems, Bae2017OnDisks} although the phenomenon is controversial and depends on the disc thermodynamics \citep{Miranda2020GapsTransport}. On the other hand, it is also possible for compact planetary systems to be responsible for the formation of single gaps morphologically degenerate with the substructure that may be generated by a single more massive planet. Even if, in some cases, a few distinguishing imprints have been found \citep{Garrido-Deutelmoser2022SubstructuresSystems} our tool was not trained on multiple systems and thus is not able to discern these scenarios.

\emph{DBNets} does not assume a precise one-to-one correspondence between planets and gaps as in our training dataset many discs present multiple substructures. Instead, at this stage, we leave to the final user the burden of interpreting in this regard the results provided based on additional information available on the specific object under exam or personal considerations.

We observed in our hydrodynamical simulations that even when multiple gaps are present at least one of them is centred on the planet's position. Therefore, when applying our tool to actual observations we suggest the user to provide as an estimate of the planet location the centre of one of the observed gaps. For example, in a system showing two gaps, one may apply the tool twice providing the location of a different gap each time. This is the approach we used in Sect. \ref{sec:real} for those observed discs showing multiple gaps. Then the two results may be considered both correct and associated to the mass of two different existing planets sufficiently distant to not influence each other substructure or, instead, one may claim that only one of the two proposed planets is actually present and it is opening both gaps. In Sect. \ref{sec:real} we provide a planet mass estimate for each gap and interpret our results according to the former interpretation. However, once one has the mass estimates from \emph{DBNets} this issue can be addressed running detailed simulations with multiple planets.

\subsubsection{Neglected physical phenomena}
\label{sec:negphe}
Like any data-driven method, our tool is strongly shaped by the synthetic observations on which we trained the CNNs. Because we used disc models rather than real data our results are all dependent on the model assumptions. For example, we assumed a viscously accreting disc which is only one of the possible paradigms proposed to explain the accretion disc evolution. Furthermore, we greatly simplified our models to reduce the number of independent physical parameters and the computational complexity of the simulations thus allowing a denser representation of the parameter space. However, to do so, we neglected several phenomena that can influence the morphology of the observed substructures.
Among these, planet migration is one of the most interesting. \cite{Nazari2019RevealingObservations}, for example, found several variations of the shape and location of planet-induced gaps when the planet was allowed to migrate, with important differences depending on the migration rate. We also neglected in our simulations the dust feedback on the gas which is known to affect the disc's morphology \citep{Dipierro2018GasBack-reaction}. The self-gravity perturbation to the star's gravitational potential can also influence the disc kinematics and thus the forming and settling of substructures \citep{Zhang2020ThePlanets}. Finally, an important aspect of our simulations that we greatly simplified is gas thermodynamics. We assumed a vertically isothermal disc maintaining the same thermal structure during the whole simulation. However, the interaction of the planet with the disc's gas could be treated more realistically including in the simulation also the energy equation. We also used two-dimensional simulations here but, especially when relaxing the local isothermal assumption, the three-dimensional structure may become important.

Our generation of synthetic observations from the dust density maps returned by the hydrodynamical simulations also presents several shortcomings. The more important lies in our treatment of the dust grain size. During the hydrodynamical simulations we fixed the Stokes number over the entire disc and then computed, using its value and the initial gas surface density, the dust grain size. This approach is affected by two problems. First, we obtain dust grains of sizes that vary across the disc with $r$. Second, at each fixed $r$ there is dust with only one grain size while, in actual discs, dust is composed of a population of grains with varying dimensions. Although multi-grain simulations would help obtain more accurate results, we can assume that our setup is a sufficient approximation for localized gaps where the dust and gas properties do not vary significantly across them and thus the morphology is primarily determined by a single Stokes number. Additionally, because we explored values $St\ll1$ we did not expect abrupt differences in the dust dynamics for little variations of its Stokes number.

\section{Conclusions}
\label{sec:conclusions}
\subsection{Achievements}
Motivated by the limitations of state-of-the-art solutions for planet detection and characterization in protoplanetary discs, we developed \emph{DBNets}, a deep-learning tool based on CNNs that exploits observations of the dust continuum emission for mass inference of gap-opening planets.
To construct this tool, we generated a large dataset of synthetic observations of protoplanetary discs and exploited it to train an ensemble of CNNs.  Throughout the development of this tool, we prioritized the study of a reliable uncertainty quantification method and a deep analysis of factors that could lead to incorrect planet mass estimates. Our goal was to provide the final user with all the material needed to interpret the planet mass estimates provided by our tool and decide on their significance.

On the test set, we achieved a standard deviation of our predictions from the target values of approximately 0.13 dex with some differences across the explored range of planetary masses. We found our tool to be most reliable for planets between $10\text M_\oplus$ and $6 M_J$.  The mean relative error on the planet mass is 23\%.

Even if the results might not be accurate on single discs, we showed, statistically, that their error is well represented by the estimated probability density functions and thus, thanks to the uncertainty quantification method developed, the use of this tool may be beneficial in the study of populations of discs. Furthermore, \emph{DBNets} could also be used in support of an analysis based on fine-tuned simulations, suggesting the more plausible planet masses to explore. In addition to its ability to represent the prediction error on the test set well, we showed that the estimated uncertainty noticeably increases when the input provided is inhomogeneous with the simulations used for training. We were able to test and quantify this phenomenon in order to provide a threshold, of 0.25 dex, prescribing that all \emph{DBNets} results with a larger uncertainty have to be rejected or, at least, considered very cautiously. We determined the conditions under which our tool is most reliable and found that in most cases we are able to determine when they are not met, observing that the estimated uncertainty increases above the threshold on which we agreed.

We also observed the importance, in the use
of the ensembling method, of constructing CNNs that are singularly able to provide an
uncertainty estimation. Solely relying on the variability of the predictions among
the ensemble’s individuals is not sufficient to correctly estimate the uncertainty.

With respect to previous works that adopted deep learning techniques for mass inference in discs \citep{Auddy2020ADisks, Auddy2021DPNNet-2.0Gaps, Zhang2022PGNets:Discs, Auddy2022UsingDisks}, we report several improvements.
\begin{itemize}
    \item We obtained comparable, or slightly better, results.  \cite{Zhang2022PGNets:Discs} obtained a standard deviation of their predictions from the actual mass values of 0.16 dex compared to our 0.13~dex; whereas \cite{Auddy2021DPNNet-2.0Gaps}, using only the disc density maps as input to their tool, achieved an r2-score of 96\% compared to ours of 97\%.
    \item We slightly extended the scope of the tool to wider ranges of disc's and planet's properties.
\end{itemize}
Most importantly, to make our tool more informative and reliable, we did the following:
\begin{itemize}
    \item We developed an uncertainty quantification pipeline that takes into account both the physical degeneracies in the planet-induced substructures and the uncertainties introduced by the deep learning methods adopted. It is important to note that while \cite{Auddy2022UsingDisks} are also able to provide uncertainties to their estimates, we used a different approach and their tool does not exploit the entire disc observations as it only takes gap's and disc's properties as input.
    \item We provided a rejection criterion of the results exceeding a prescribed uncertainty.
    \item We tested our tool against out-of-distribution data to outline under which conditions it can be reliably applied.
\end{itemize}

Compared to the empirical relations that we derived on our same dataset, in line with those proposed in the literature, our tool obtains higher scores in all the metrics that we evaluated and maintains the same efficiency and velocity of applicability of the empirical formulae. However, our tool has the significant advantage of being able to exploit the entire observation and of not requiring the manual measurement of substructures' properties that could introduce errors and biases linked with their definition.

Finally, we applied \emph{DBNets} to 33 discs with visible substructures in the dust continuum. We propose a population of 48 planets to be emerging from these substructures, providing the most comprehensive analysis of their masses relying on a common method and able to take into account several physical degeneracies. We found super-Jupiter planets to be a small fraction of our sample (32\%) despite being more sensitive to these objects as they would carve more prominent substructures. We thus conclude, in agreement with previous works and failed detections, that the number of super-Jupiter planets in protoplanetary discs at 10-100 au cannot be significant.

\subsection{Limitations and future perspectives}
We outlined in Sect. \ref{sec:caveats} the main limitations of this tool due to the simplifying assumptions that we adopted to generate -- within reasonable computational costs -- synthetic data able to densely represent a wide region of the parameter space. At this stage, our assumptions are mostly aligned with other state-of-the-art methods used for planet mass inference. In future works, we will study the effect of the physical processes neglected in this work, include them in our training dataset, and provide a retrained and improved version of our tool.

Additionally, a significant improvement could be obtained by extending the tool's analysis
to discs’ observations in different tracers. The use of kinematical data, in particular, will
be the major step. The intrinsic differences in the planetary signatures in these data will enable a fruitful integration of the two to break some degeneracies and
increase the accuracy and statistical significance of the planet mass estimates.

\section{\emph{DBNets}: Tool availability and instructions}
\emph{DBNets} is publicly available both as web-based interface\footnote{ \href{http://dbnets.fisica.unimi.it}{dbnets.fisica.unimi.it}} and as a python library\footnote{\href{https://github.com/dust-busters/DBNets}{https://github.com/dust-busters/DBNets}}. The tool is very easy and fast to use. We strongly encourage the inclusion of \emph{DBNets} in the analysis pipeline of any continuum observation, showing substructures to evaluate the hypothesis of a planetary origin. \emph{DBNets} requires a disc observation of the dust continuum, the disc geometrical properties (inclination, position angle, and location of the disc centre within the image), and a guess on the alleged planet location either provided as the angular separation from the disc centre or in physical units with the disc distance. The tool thus returns a probability density function for $\log M_p$ and provides utility functions for its interpretation. We suggest the use of the python library for complete access to all \emph{DBNets} functionalities, such as the possibility to automatically apply the tool on several discs. Comprehensive documentation with commented examples is available via the same links.


\begin{acknowledgements}
We thank the anonymous referee for the detailed report and for the insightful comments that improved the manuscript. 
Computational resources have been provided by INDACO Core facility, which is a project of High Performance Computing
at the Università degli Studi di Milano (\href{https://www.unimi.it}{https://www.unimi.it}).
This work has been supported by Fondazione Cariplo, grant n° 2022-1217, from the European Union’s Horizon Europe Research \& Innovation Programme under the
Marie Sklodowska-Curie grant agreement No. 823823 (DUSTBUSTERS) and from the European Research Council (ERC) under grant agreement no. 101039651 (DiscEvol). Views and opinions expressed are however those of the author(s) only, and do not necessarily reflect those of the European Union or the European Research Council Executive Agency. Neither the European Union nor the granting authority can be held responsible for them.

\end{acknowledgements}

\bibliographystyle{aa}
\bibliography{references,inprep}

\begin{appendix}
    
\section{Resolution of protoplanetary discs continuum observations}
\label{app:resreal}

\begin{figure}[h]
    \centering
    \vspace{1cm}
\includegraphics{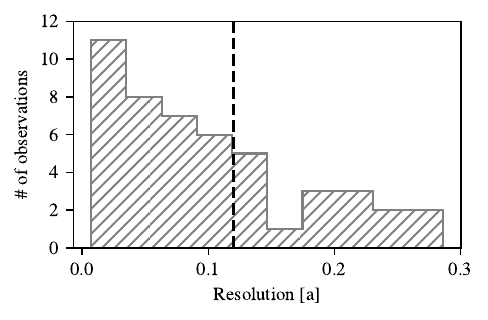}
\caption{ Distribution of the resolution of dust continuum observations listed in Table \ref{tab:prop}. The black dashed line marks the simulated resolution of the synthetic data generated and used in this work.}
    \label{fig:resreal}
    \vspace{0.2cm}
\end{figure}
\vspace{0.3cm}

\begin{figure}[h]
    \centering
\includegraphics{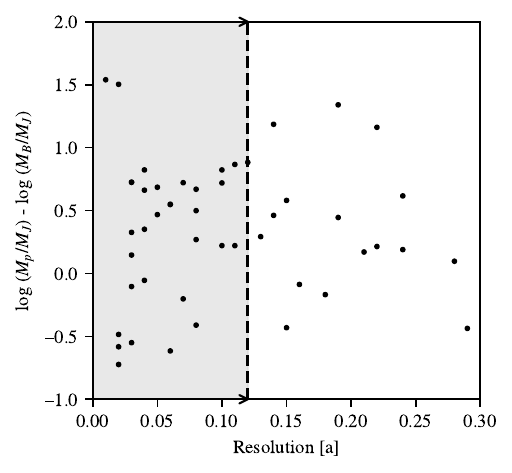}
\caption{ Deviations, for the actual observations considered in Sect. \ref{sec:real}, of \emph{DBNets} mass estimates ($M_p$) from the literature values ($M_B$) reported in \cite{Bae2023StructuredDisks}, as a function of the resolution of the images. Note that all the observations in the shaded area have been additionally convolved with an appropriate Gaussian beam to match the resolution of the training data ($0.12a$, black dashed line) as explained in Sect. \ref{sec:real}. }
    \label{fig:deviationres}
\end{figure}

\vspace{0.7cm}

\begin{figure}[h]
\vspace{1cm}
    \centering
    
\includegraphics{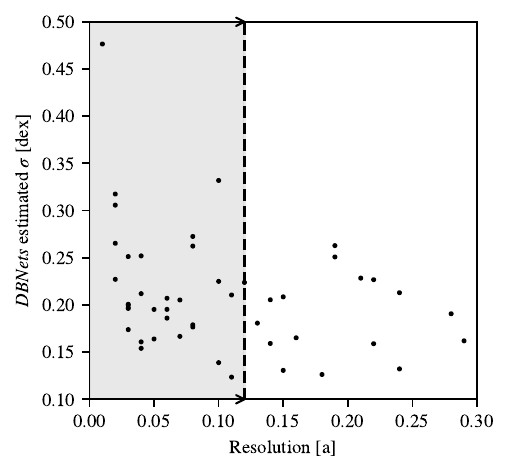}
\caption{ \emph{DBNets} estimated uncertainty $\sigma$ for the actual observations considered in Sect. \ref{sec:real} as a function of the images resolution. Note that all the observations in the shaded area have been additionally convolved with an appropriate Gaussian beam to match the resolution of the training data ($0.12a$, black dashed line) as explained in Sect. \ref{sec:real}.}
    \label{fig:sigmares}
    \vspace{0.5cm}
\end{figure}

The histogram in Fig. \ref{fig:resreal} presents the distribution of continuum observations resolutions over the sample of discs considered in Sect.~\ref{sec:real} (see Table \ref{tab:prop} for a list of the selected discs). The binned resolution is, in accordance with the rest of this work, expressed as the (major) standard deviation of the observational beam Gaussian approximation, in units of the putative planet(s) radial location. Due to this rescaling, for some objects, multiple values are associated with the same observation. The black dashed line marks the simulated resolution of the synthetic data generated and used in this work.

With some tests on numerical simulations where the target planet mass is known, we found  as explained in Appendix \ref{app:ood} that our tool provides the best results when applied to observations with a resolution similar to that of the training data. For this reason, as explained in Sect. \ref{sec:real}, we additionally convolved each better-resolved image with an appropriate Gaussian beam to match the resolution of the training data. We checked the results obtained with \emph{DBNets} in all the other cases when the image resolution is worse but we did not find any systematic difference in our estimates for this group of observations with respect to the other results. This can be observed in Figs. \ref{fig:deviationres} and \ref{fig:sigmares} where we show,  for the actual observations considered in Sect. 4, respectively the deviations
of \emph{DBNets} mass estimates from the literature values reported in \cite{Bae2023StructuredDisks} and \emph{DBNets} estimated uncertainties. Both are shown as a function of the images' resolution.

\section{Peers versus experts mode}
\label{sec:peersvsphysics}

\begin{figure*}[h!]
\vspace{2cm}
    \centering
   \includegraphics{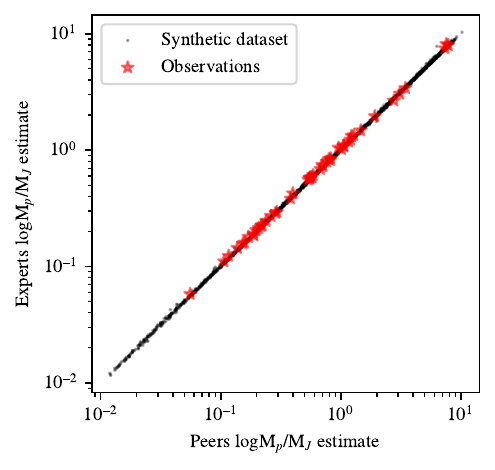}
   \includegraphics{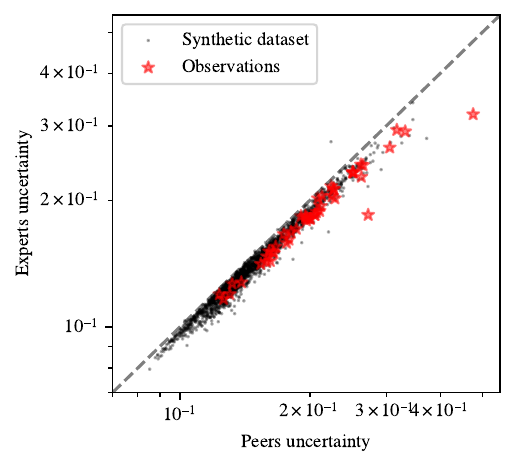}
   \vspace{1cm}
    \caption{Comparison between the peers and experts ensembling modes. Left panel: Comparison of the mass estimates obtained with the two ensembling
modes for the test set (black points) and actual observations (red stars). Right panel: Comparison of the estimated uncertainties of \emph{DBNets} mass measures obtained with the two ensembling
modes for the test set (black points) and actual observations (red stars). The dashed grey line marks the identity relation.}
    \label{fig:pe1}
\vspace{1.55cm}
\end{figure*}

In Sect. \ref{sec:ensemble} we presented two different methods for determining the final estimate of the ensemble from the outputs of the single neural networks which we called experts and peers mode. These two methods are expected to be significantly different when the uncertainties estimated by the neural networks in the ensemble are very different from each other as the experts mode would skew the results in favour of the CNN estimating the lower uncertainty. We compared these two ensembling modes on the test set observing little difference between them as shown in Fig. \ref{fig:pe1}. The idea at the base of the
experts mode is the possibility that different neural networks might specialize, during
training, on particular subclasses of disc observations based on peculiar morphologies or
physical parameters. The CNN would then learn to assign lower uncertainties to the mass
estimates in its domain of specialization. However, we observed that on the test set,
which is a group of observations that covers exhaustively the domain of the training set,
the results obtained are always very similar. We thus interpret these results as evidence that this specialization is not happening and decided to opt for the peers mode as the
default ensembling method. We note that while on the test set the two methods provide
almost identical results, the same might not be always true when the tool is generalized to other data. We observe for example in the right panel of Fig. \ref{fig:pe1} that in two instances, that is, the gap at 17 au in HD169142 and the gap at 81 au in TW Hya, the experts approach leads to a significant underestimation of the proposed planet's mass uncertainty due to an overconfident estimate of one member of the ensemble. We did not observe any peculiar feature of these images that could justify this result but we note that, in both cases, the estimated  uncertainty  (with the peers method) overcomes the established  threshold for rejection. The peers method is the 
conservative \newpage \noindent choice because relies less on the single estimates of the ensemble’s individuals
that might incorrectly return a small uncertainty compromising the results.

\section{Learning curves}
\label{app:learncurv}

\begin{figure*}[h!]

    \centering
    \scalebox{0.93}{\includegraphics{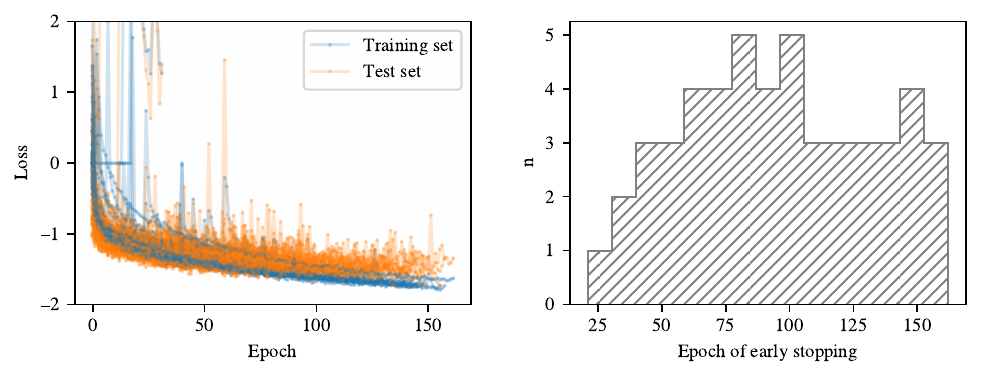}}
    \caption{Training of the CNNs ensemble. Left panel: Training curves (loss function at each training epoch) of all the 50 CNNs of the final ensemble; for the blue lines the loss function is computed on the training set while the orange lines represent the results obtained on the test set. Right panel: Distribution of the number of epochs after which, monitoring the value of the loss function on the test set, the CNN training was early stopped.
    }
    \label{fig:training}
\end{figure*}
Figure \ref{fig:training} shows the CNNs learning curves: the loss function computed at each epoch of training, of all the CNNs of the ensemble.
The number of epochs, after which the training of each CNN in the ensemble is early stopped, varies widely between 25 and 150 as shown in the right panel of Fig. \ref{fig:training}. The warm-up phase that precedes the actual training with the complete loss function may be the reason why the CNNs training is fast as it already shapes the networks to give the correct planet mass estimates.
We observe that the learning curves are all rather similar, at the beginning of training the loss function on both the test and training set gradually decreases until on the test set it stops improving. Early stopping prevents the models from overlearning the training dataset. We also observe in Fig. \ref{fig:training} that all the learning curves evaluated on the test set settle approximately around~the same value.

\FloatBarrier

\section{Out-of-distribution data detection, additional effects}
\label{app:ood}
Below we report more observational non-idealities for which we tested their effect on \emph{DBNets} results in addition to those presented in Sect.~\ref{sec:oodd}.

{\bfseries Observational resolution}: we simulated this observational non-ideality convolving the synthetic images with a Gaussian beam whose size is related to the observational resolution. We tried synthetic beams of different sizes and found that \emph{DBNets} works best on observations with the same resolution of the data used for training which corresponds to a beam size of $0.12 a$ (with $a$ being the planet orbital radius). 
In this case,
we observed that when the image resolution is much worse ($\gtrsim 0.3a$), the uncertainty of our mass estimates is higher than 0.25 dex.
Although the resolution can be easily checked manually, this result is important as it shows, for a different type of distribution shift, that \emph{DBNets} is able to detect out-of-distribution data.

{\bfseries Information loss due to deprojection}: because we trained our deep learning models only on face-on disc's images, before applying them to actual observations these need to be deprojected if the disc is observed from a different inclination. 
We found that the results are not affected by this loss of information as long as the inclination remains below approximately 60°. Additionally, when the disc inclination is higher the estimated uncertainty also increases above the 0.25 dex threshold. 

{\bfseries Planet position}: we evaluated how the planet mass estimate returned by \emph{DBNets} depends upon the guessed planet position used to rescale the observation. We found that for accurate planet mass estimates using our tool, the position of the gap under exam must be known with at least a 5\% accuracy. In this case, when \emph{DBNets} mass estimates become unreliable the uncertainty doesn't always increase above the 0.25 dex threshold. One could apply the tool varying the planet's location within a given range and adopt as the final mass estimate the one related to the smallest uncertainty. A more conservative approach would be to exploit the fact that \emph{DBNets} returns a probability distribution function and marginalize its dependence on the planet location given an appropriate prior.

\section{Stability of the proposed planetary systems}
\label{app:stability}

\begin{figure}[h!]
    \centering
    \scalebox{0.93}{\includegraphics{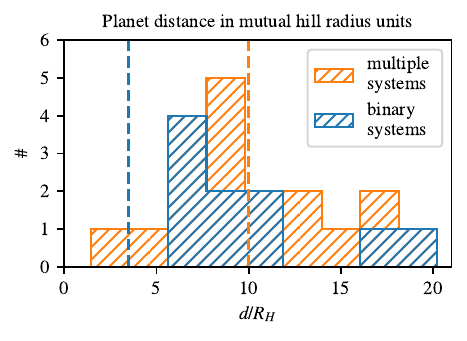}}
    \caption{Distance in units of the mutual Hill radius of each pair of planets proposed in discs' observations with multiple dust substructures. The dashed lines mark the lower limits for the system stability found by \cite{Gladman1993DynamicsPlanets} and \cite{Chambers1996TheSystems}.}
    \label{fig:stability}
\end{figure}

To address in more detail the stability issue discussed in Sect. \ref{sec:stability}, we checked the stability, according to the Hill criterion, of the planetary systems we propose in discs with multiple substructures. \cite{Gladman1993DynamicsPlanets} found that binary systems are stable if the mutual distance $(d)$ of the planets, expressed in terms of their mutual Hill radius $    R_H = \left[{(m_1+ m_2)}/{3M_\star}\right]^{1/3}\left[{(a_1+a_2)}/{2}\right]$ is greater than $2\sqrt 3$. Here $a_1$ and $a_2$ are the semi-major axes of the two planet orbits. \cite{Chambers1996TheSystems} extended this analysis investigating numerically the stability of multiple systems. Their results show that all binary subsystems whose mutual distance $d/R_H$, defined as before, is lower than 10 are unstable.
We computed, and show in Fig. \ref{fig:stability}, $d/R_H$ for each pair of planets proposed in the sample of actual discs studied in Sect. \ref{sec:real} using the planet masses that we proposed using our tool. The only five pairs of planets whose distance is below the instability threshold (in descending order of $d/R_H$) are those proposed in the inner and outer gaps of HD169142, in the two outer gaps of HL Tau, in the two outer gaps of HD163296,  in the two outer gaps of HD169142, and in the two outer gaps of DL Tau.

\end{appendix}

\end{document}